# Decoupled Water Splitting: From Basic Science to Application

Avigail Landman,[1] Gideon S. Grader[1,*] and Avner Rothschild[2,*]

[1] Department of Chemical Engineering, Technion – Israel Institute of Technology, Haifa, Israel.

[1] Department of Materials Science and Engineering, Technion – Israel Institute of Technology, Haifa, Israel.

* E-mail: grader@technion.ac.il (GSG) ; avnerrot@technion.ac.il (AR)

## Abstract

As we move away from fossil fuels towards renewable energies, green hydrogen produced by water electrolysis becomes a promising and tangible solution for storage of excess energy for power generation and grid balancing, and for the production of decarbonized fuel for transportation, heating and other applications. To compete with the price set by steam methane reforming and other carbon-based hydrogen production technologies, a true paradigm shift in water electrolysis technologies is required. In this review, we discuss disruptive decoupled water splitting schemes, in which the concurrent production of hydrogen and oxygen in close proximity to each other in conventional electrolysis is replaced by time- or space-separated hydrogen and oxygen production steps. We present the main decoupling strategies, including electrolytic and electrochemical – chemical water splitting cycles, and the redox materials that facilitate them by mediating the ion exchange between the hydrogen and oxygen evolution reactions. Decoupled water splitting offers increased flexibility and robustness and provides new opportunities for hydrogen production from renewable sources.

## Keyword list

**Introduction**

Since its discovery by Troostwijk and Diemann in 1789,[1] alkaline electrolysis (AEL) has become a mature technology, applied for large-scale hydrogen production for over a century.[2] Some years later, in the mid-20th century, proton exchange membrane electrolysis (PEMEL) was introduced as a compact alternative to AEL, replacing the alkaline aqueous electrolyte with a polymer-based proton conducting membrane.[3,4] More recently, solid oxide electrolysis (SOEL) and anion exchange membrane electrolysis (AEMEL) have been introduced. Though promising in many regards, these new technologies still need further development to reach acceptable durability and performance.[5] Throughout its various embodiments, the structure of the basic electrolyzer unit has remained essentially unchanged; Namely, it comprises of an electrolytic cell with an anode, a cathode, and a separator, where hydrogen and oxygen are produced concurrently in the same cell. Thus, the two reactions are *coupled* in both time and space.

Currently, the main end-uses of hydrogen are in ammonia production and oil refining, and water electrolysis accounts for approximately 2% of the global hydrogen production.[6] However, as our energy landscape shifts away from fossil-based fuels and towards renewable energies,[7] hydrogen is predicted to play an important role as an energy vector, with water electrolysis being the most promising technology to produce green hydrogen using renewable power sources.[8,9] In this respect, the coupling of the hydrogen and oxygen evolution reactions in conventional electrolyzers presents some critical challenges.

The first issue is that of safety under partial-load operation conditions, which are inherent to intermittent power sources such as solar and wind. At low operation currents, the ratio between the rate of gas permeation through the separator and the rate of gas evolution at the electrodes becomes significant, leading, potentially, to hazardous $H_2/O_2$ mixtures within the cell.[2,10] In addition to posing a potential safety hazard, this mixing also reduces the electrolyzer efficiency as hydrogen and oxygen recombine to form water. Finally, it contributes to membrane degradation by reactive oxygen species that form in $H_2/O_2$ mixtures in the presence of catalysts.[11] Although this issue is less severe in PEM electrolyzers, it becomes more significant for thinner membranes and under high pressure operation.[3,12] Therefore, the integration of alkaline and PEM electrolyzers with intermittent power sources is a challenge that must be addressed.

The second issue is that of the cost of hydrogen. The U.S. Department of Energy cost target for hydrogen produced by electrolysis is $ 2-2.3 per kg $H_2$ by the year 2020.[13] This target is currently beyond reach by existing electrolysis systems. A recent techno-economic analysis of green hydrogen production by photovoltaic (PV) powered electrolysis reports an expected production cost of more than $ 6 per kg $H_2$.[14] A sensitivity analysis suggests that the electrolyzer efficiency has the largest impact on this cost, and the authors claim that *"The development of a truly disruptive electrolysis technology is required"* to attain cost-competitive production of green hydrogen from renewable sources.

The third issue concerns the pressure at which hydrogen can be produced. In most applications, hydrogen is supplied in trucks that distribute it at 200 bar. At refueling stations of busses and cars, hydrogen is typically compressed to 350 and 700 bar, respectively. However, commercial alkaline and PEM electrolyzers typically produce hydrogen at 10 and 30 bar, respectively. This limitation is due to increased hydrogen crossover rate through the separator at high pressures. To raise the pressure to the desired level, costly and energy intensive compressors must be used.

Finally, the conventional electrolysis configuration gives rise to additional challenges in photoelectrochemical (PEC) solar water splitting, wherein at least one of the electrodes is photoactive (photoanode or photocathode) and must therefore be placed in the solar field and exposed directly to sunlight, similarly to photovoltaic (PV) cells. This requirement leads to distributed production of hydrogen in a large number of PEC cells, resulting in a severe challenge of hydrogen gas collection and transport to a central storage and distribution facility, as discussed previously.

It is in light of these challenges that *decoupled* water splitting was first proposed by Symes and Cronin in 2013.[15] This work introduced a disruptive water splitting concept wherein instead of the oxygen evolution reaction (OER) taking place concurrently with the hydrogen evolution reaction (HER), as in conventional water electrolysis, a soluble redox mediator (phosphomolybdic acid) is oxidized at an auxiliary electrode while hydrogen is evolved at the cathode, without concurrent oxygen generation. Then, the oxidized redox mediator is reduced at the auxiliary electrode while oxygen is evolved at the anode, without concurrent hydrogen generation, to complete the water splitting reaction. Hence, decoupling of the HER and OER is achieved in both time and space. Following Symes and Cronin's pioneering work,[15] other studies on soluble redox mediators for decoupled water splitting in various operation schemes have emerged.[16–21] An interesting modification was reported by Rausch *et al.* in 2014,[22] proposing an electrochemical - chemical cycle wherein after the oxygen evolution step, the reduced redox mediator (silicotungstic acid) undergoes a spontaneous chemical oxidation back to its initial state in the presence of a suitable catalyst while releasing hydrogen.

The use of solid redox mediators (instead of soluble ones) was first proposed by Rothschild *et al.* in a patent application filed in 2014,[23] followed by publications by Chen *et al.* (2016) and by Landman *et al.* (2017),[24,25] using nickel hydroxide/oxyhydroxide ($Ni(OH)_2$/NiOOH) auxiliary electrodes which are commonly used in secondary alkaline batteries. Soon thereafter, other solid redox mediators were reported in both alkaline and acidic media.[26–29] A few years later, in 2019, Dotan *et al.* demonstrated an electrochemical – chemical cycle using a nickel hydroxide anode, wherein the anode is first oxidized electrochemically during the hydrogen evolution step, and is then regenerated back to its initial state by a thermally-activation chemical reaction with water, evolving oxygen and completing the water splitting reaction.[30] Another unique and fundamentally different approach to decoupled water splitting is pursued by the French start-up company Ergosup.[31] Ergosup's technology is based on the electrochemical – chemical cycle of zinc (Zn) electrodeposition and dissolution, and enables high-pressure hydrogen production.[32]

Another interesting application of decoupled water splitting is in PEC solar water splitting, as a disruptive strategy to overcome the challenge of hydrogen gas collection and transport from distributed PEC cells, as discussed above. In 2016, Bloor *et al.*[33] reported the first decoupled PEC water splitting system using a soluble redox mediator and a tungstic oxide ($WO_3$) photoanode in an acidic electrolyte. The use of a redox mediator allows the separation of the PEC cell into two separate cells, a PEC oxygen evolution cell and an electrochemical hydrogen evolution cell. A year later, Landman *et al.*[25] proposed another concept for decoupled PEC water splitting, using solid redox electrodes and a hematite ($Fe_2O_3$) photoanode in an alkaline electrolyte. This concept was recently realized and demonstrated at a benchtop scale.[34] Finally, a photocatalytic scheme for decoupled water splitting was proposed by Li *et al.* in 2017, using a soluble redox mediator and a bismuth vanadate ($BiVO_4$) photocatalyst.

This work reviews the various decoupling strategies, discussing their advantages, disadvantages, and future prospects.

**Electrolytic schemes for decoupled water splitting**

The concept of decoupled water splitting in an electrolytic scheme is principally based on the integration of a redox mediator into the water splitting process. The redox mediator mediates the ion exchange between the hydrogen evolving cathode and the oxygen evolving anode by undergoing reversible oxidation and reduction reactions. These reactions occur at another set of electrodes, besides the cathode and anode that facilitate the HER and OER, respectively. We refer to the hydrogen evolving cathode and the oxygen evolving anode as the *primary electrodes*, whereas the electrodes that facilitate the redox reactions of the mediator are referred to as *auxiliary electrodes*. Integration of the redox mediator enables to separate the electrolytic cell into two cells that generate hydrogen and oxygen separately, rather than in the same cell as in conventional electrolysis. In the hydrogen cell, the redox mediator undergoes oxidation at an auxiliary anode while hydrogen is generated at the primary cathode. Similarly, in the oxygen cell the redox mediator undergoes reduction at an auxiliary cathode while oxygen is generated at the primary anode. The ability of the redox mediator to undergo reversible redox cycles without parasitic reactions such as oxygen or hydrogen evolution is paramount to decoupled water splitting without $H_2/O_2$ mixing.

Since the redox mediators in the oxygen and hydrogen cells have finite charge capacities, the operation must be stopped when their full operational capacity has been reached, and they must be regenerated to continue operation. The redox mediators can be regenerated by subjecting them to a reverse polarization, such that the mediator that was oxidized in the first step will be reduced in the next step, alongside oxygen evolution at the primary anode, and the previously reduced mediator will be oxidized in the next step, alongside hydrogen evolution at the primary cathode. Alternatively, regeneration of the redox mediator can be carried out by swapping the reduced mediator from the oxygen cell with the oxidized mediator from the hydrogen cell.

**Electrolytic water splitting with soluble redox mediators**

Symes and Cronin were the first to introduce the concept of decoupled water splitting, using phosphomolybdic acid (PMA, $(H_3O^+)[H_2PMo_{12}O_{40}]^-/(H_3O^+)[H_4PMo_{12}O_{40}]^-$) as an electron-coupled proton buffer (ECPB).[15] The $(H_3O^+)[H_2PMo_{12}O_{40}]^-/(H_3O^+)[H_4PMo_{12}O_{40}]^-$ couple serves as a soluble redox mediator that undergoes reversible oxidation and reduction reactions by taking up or releasing the protons and electrons that are produced in the OER and consumed in the HER, respectively. As discussed above, the redox mediator undergoes oxidation at an auxiliary anode and reduction at an auxiliary cathode, while the HER and OER take place at the primary cathode and anode, respectively. Thus, to avoid unintentional reduction of the soluble redox mediator at the primary cathode during the hydrogen evolution step, or its unintentional oxidation at the primary anode during the oxygen evolution step, the cell is divided into two compartments, the primary electrode compartment and the auxiliary electrode compartment, using a proton exchange membrane (PEM).

The water splitting process is divided into two steps, as illustrated in Figure 1. The first step (Step I) is an oxygen generation step, during which the OER takes place at the primary anode while reduction of the PMA takes place at the counter auxiliary electrode:

$$\text{Anode: } H_2O \rightarrow \tfrac{1}{2}O_2 + 2H^+ + 2e^-, \ E^0 = 1.23 \ V_{RHE} \quad \text{(Rxn 1)}$$

$$\text{Auxiliary electrode: } [H_2PMo_{12}O_{40}]^- + 2e^- + 2H^+ \rightarrow [H_4PMo_{12}O_{40}]^-, \ E^0 \cong 0.65 \ V_{RHE} \quad \text{(Rxn 2)}.$$

$E^0$ is the standard potential of the respective reaction and $V_{RHE}$ is volts against the reversible hydrogen electrode. The protons evolved in the OER (Rxn 1) travel from the primary anode through the anolyte (1M $H_2PO_4$ in water) and through the catholyte (0.5M 50%:50% mix of the reduced and oxidized PMA in water) to the auxiliary electrode, where they are taken up by the PMA reduction reaction (Rxn 2), alongside the electrons that travel through the external electrical circuit. Step I proceeds until the PMA is reduced. Then, in the next step (Step II), the current polarity is reversed, and the reduced PMA is re-oxidized at the auxiliary electrode according to Rxn 2 (reversed) while the HER takes place at the primary cathode:

$$\text{Auxiliary electrode: } [H_4PMo_{12}O_{40}]^- \rightarrow [H_2PMo_{12}O_{40}]^- + 2e^- + 2H^+, \ E^0 \cong 0.65 \ V_{RHE} \quad \text{(Rxn 2, reversed)}$$

$$\text{Cathode: } 2H^+ + 2e^- \rightarrow H_2, \ E^0 = 0 \ V_{RHE} \quad \text{(Rxn 3)}.$$

The protons evolved in the PMA oxidation reaction (Rxn 2, reversed) travel to the cathode where they are reduced by the HER (Rxn 3), forming hydrogen at the cathode. All in all, the sum of Rxns 1 through 3 that take place in Step I and Step II yields the overall water splitting reaction:

$$H_2O \rightarrow H_2 + \tfrac{1}{2}O_2 \quad \text{(Rxn 4)}.$$

At the end of Step II, the PMA that has been reduced in Step I is re-oxidized back to its initial state, and the entire process can be repeated for another cycle. The ability of the PMA to repeatedly undergo reversible redox cycles by taking-up and releasing protons (and electrons) enables operation without pH oscillations. Thereby, it serves as a proton-coupled electron buffer (ECPB) that mediates the proton exchange between the primary electrodes that facilitate the water splitting reactions (OER and HER).

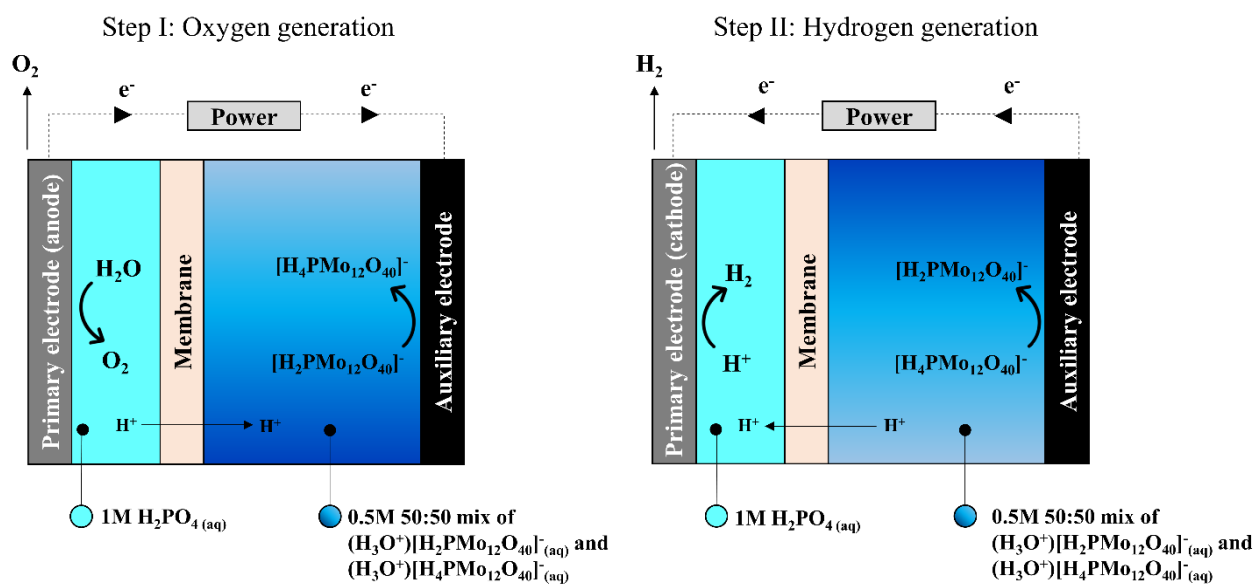

*Figure 1: Schematic illustration of the decoupled water splitting scheme proposed by Symes and Cronin using PMA ( $(H_3O^+)[H_2PMo_{12}O_{40}]^- / (H_3O^+)[H_4PMo_{12}O_{40}]^-$ ) as a soluble redox mediator.[15] In the oxygen generation step (Step I, left), the redox mediator ( $[H_2PMo_{12}O_{40}]^-$ ) is reduced at the auxiliary electrode while oxygen is evolved at the primary electrode (serving as an anode). Then, in the hydrogen generation step (Step II, right), the reduced redox mediator ( $[H_4PMo_{12}O_{40}]^-$ ) is oxidized at the auxiliary electrode back to its initial state ( $[H_2PMo_{12}O_{40}]^-$ ) while hydrogen is evolved at the primary electrode (serving as a cathode).*

In order to mediate the proton exchange between the HER and the OER, the redox mediator must have a redox potential within the water stability region, *i.e.*, between the HER and OER standard potentials.

PMA has several redox waves within this region, as shown in the cyclic voltammogram (CV) in Figure 2. Specifically, the redox wave corresponding to the transition between its oxidized state ($[H_2PMo_{12}O_{40}]^-$) and its two-electron-reduced state ($[H_4PMo_{12}O_{40}]^-$), the two states that participate in the water splitting process (Rxn 2, forward and reversed), is centered around ~0.65 $V_{RHE}$, well above the HER standard potential and below the OER standard potential. This ensures that no hydrogen evolves at the auxiliary electrode during the oxygen generation step, and no oxygen evolves at the auxiliary electrode during the hydrogen generation step. PMA has several additional attributes which make it an attractive soluble redox mediator; Namely, it is electrochemically stable under cycling (when not exposed to air), it has high solubility in water (1.62 g ml$^{-1}$ at 25°C), and it enables complete separation of the $H_2$ and $O_2$ products with 100% Faradaic efficiency. However, due to the corrosive acidic environment (pH 0.3) required for the PMA's stability, this system requires precious metal catalysts, similarly to PEM electrolysis. In Symes and Cronin's pioneering work,[15] Pt was used for the anode, cathode and the auxiliary electrode.

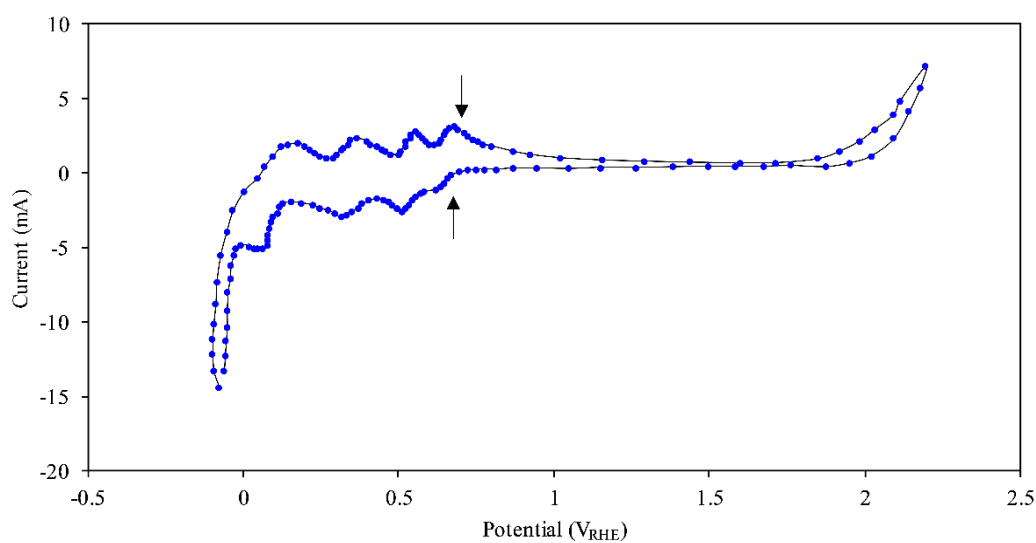

*Figure 2: Cyclic voltammogram measured with a Pt disc working electrode in an aqueous solution of 0.5M phosphomolybdic acid (PMA). Plot reconstructed according to data from* [15].

In this seminal work,[15] Symes and Cronin laid the basic material selection criteria of redox mediators for decoupled water splitting. Expanding upon their criteria, the following guidelines are proposed for the selection of suitable redox mediators for decoupled water splitting:

(i) **Stability:** The redox mediator must be chemically stable in an aqueous electrolyte under the operation conditions of the electrolytic cell.

(ii) **Reversibility:** The redox mediator should have a reversible redox reaction that exchanges electrons and ions with the auxiliary electrode and the electrolyte, respectively. No other reactions should be facilitated by the redox mediator within the entire operation range of the electrolytic cell.

(iii) **Redox potential:** The reversible potential of the redox reaction from guideline (ii) should be within the water stability range, *i.e.*, above the HER standard potential and below the OER standard potential. Otherwise, oxygen and hydrogen may evolve during the redox mediator's oxidation and reduction reactions, respectively, leading to $H_2/O_2$ mixing. In practice, this range may be extended to the effective water stability range, *i.e.*, above the onset potential for hydrogen evolution at the primary cathode and below the onset potential for oxygen evolution at the primary anode.

(iv) **Low overpotential:** The redox mediator's oxidation and reduction overpotentials should be low to minimize the applied voltage and the resultant power losses that are incurred by the decoupling process.

(v) **Cyclability:** The redox mediator should withstand many redox cycles without performance deterioration during the system's lifetime.

(vi) **Compatibility:** The redox mediator should be compatible with the other cell components (electrodes, membrane, etc.), to prevent their degradation. Specifically, it should not release any substance that may poison the primary electrodes or foul the membrane.

(vii) **Capacity:** The redox mediator should have a high charge capacity per unit volume and/or weight.

(viii) **pH buffering:** The mediator's redox reactions should not affect the pH of the electrolyte to avoid substantial pH gradients during operation.

(ix) **Faradaic efficiency:** The Faradaic efficiency of the redox mediator's oxidation and reduction reactions should be 100%, *i.e.*, no reactions other than oxidation/reduction of the mediator should take place, and no side-products should be produced.

(x) **Cost and availability:** For industrial systems, the redox mediator should comprise of Earth abundant elements, be non-toxic and environmentally benign, and cheap.

For soluble redox mediators, guideline (vii) requires high molecular charge capacity and high solubility in water under the operation conditions of the electrolytic cell. Furthermore, in acidic media, guideline (viii) requires that the only counter ion in the mediator's cycling redox reactions should be a proton ($H^+$).

Guideline (iv) relates to the losses that are incurred by the addition of another component, the redox mediator, to the electrolytic cell. Due to the overpotential losses of the oxidation and reduction reactions of the redox mediator, the total voltage that must be applied for the overall decoupled process ($V_{decoupled}$), *i.e.*, the sum of the individual cell voltages of the hydrogen and oxygen cells, ($V_{hydrogen\ cell} + V_{oxygen\ cell}$), is greater than the corresponding voltage for conventional (coupled) water splitting with the same primary electrodes ($V_{coupled}$):

$$\begin{aligned}
V_{decoupled} &= V_{hydrogen\ cell} + V_{oxygen\ cell} = (E_{OER} - E_{red}) + (E_{ox} - E_{HER}) + \sum iR = \\
&= [(1.23 V_{RHE} + \eta_{OER}) - (E^0_{redox} - \eta_{red})] + [(E^0_{redox} + \eta_{ox}) - (0 V_{RHE} - \eta_{HER})] + \left( \sum_{coupled} iR + \Delta \sum iR \right) = \\
&= [1.23 V + \eta_{OER} + \eta_{HER} + \sum_{coupled} iR] + \eta_{ox} + \eta_{red} + \Delta \sum iR = \\
&= V_{coupled} + \eta_{ox} + \eta_{red} + \Delta \sum iR
\end{aligned}$$
(Eq 1)

where $E_{OER}$ and $E_{HER}$ are the potentials of the oxygen and hydrogen evolution reactions under polarization, respectively, $E_{ox}$ and $E_{red}$ are the oxidation and reduction potentials of the redox mediator under polarization, respectively, $\sum iR$ is the sum of all the Ohmic losses (including the series resistances of the electrodes, electrolyte, membrane and electrical contacts), 1.23 $V_{RHE}$, 0 $V_{RHE}$ and $E^0_{redox}$ are the standard potentials of the OER, HER and the mediator's redox reaction, respectively, and $\eta_{OER}$, $\eta_{HER}$, $\eta_{ox}$, and $\eta_{red}$ are the overpotentials of the OER, HER, and the redox mediator oxidation and reduction reactions, respectively. $\Delta\sum iR$ is the difference in the Ohmic losses between the decoupled and coupled processes. Consequently, the total voltage that must be applied for the decoupled water splitting process is necessarily greater than that of an equivalent coupled process ($V_{coupled}$) due to the polarization overpotentials of the redox mediator's oxidation and reduction reactions.

Figure 3 shows the current density – voltage curves taken with and without PMA. Using 0.5M PMA (a 50%:50% mix of $(H_3O^+)[H_2PMo_{12}O_{40}]^-$ and $(H_3O^+)[H_4PMo_{12}O_{40}]^-$) and 1M $H_3PO_4$ as the electrolytes in the two cell compartments, a Nafion membrane and Pt electrodes operated at a current density of 100 mA/cm$^2$, Symes and Cronin demonstrated decoupled water splitting operation with applied voltages of approximately 1.23 and 1.6 V for the hydrogen and oxygen cells, respectively, and a total voltage ($V_{decoupled}$) of 2.83 V. The equivalent coupled water splitting voltage ($V_{coupled}$), using the same Pt electrodes in one cell with 1M $H_2PO_4$ aqueous electrolyte, was 2.25 V, corresponding to a decoupling loss of 1 - 2.25V/2.83V = 20%.

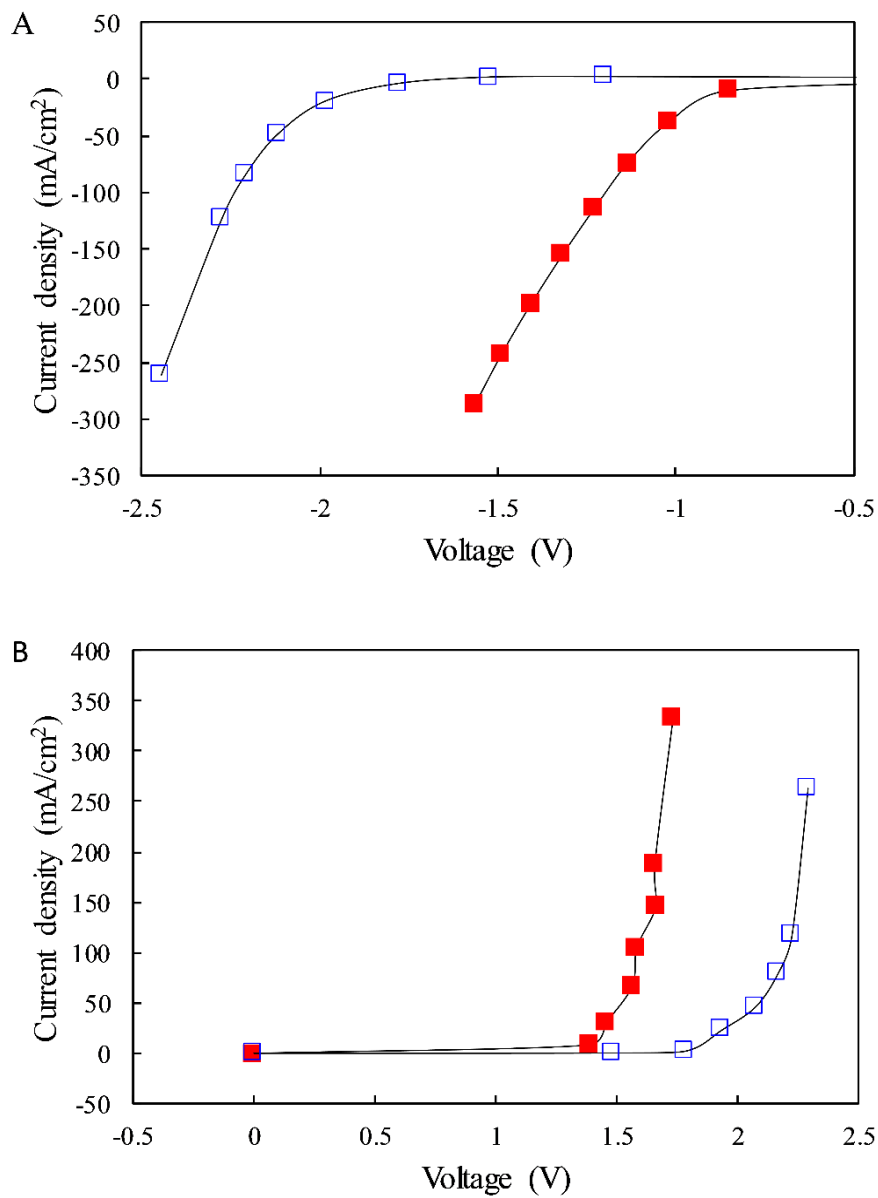

*Figure 3: Current density – voltage curves with (red) and without (blue) PMA in the hydrogen cell (A) and the oxygen cell (B). Pt discs were used for the primary electrodes (for the HER in (A) and the OER in (B)), and large-area Pt mesh electrodes were used as the auxiliary electrodes. The primary electrodes were dipped in 1M $H_3PO_4$ aqueous electrolyte and the auxiliary electrodes were dipped either in 1M $H_3PO_4$ aqueous electrolyte (blue) or in 0.5M PMA aqueous electrolyte with 50%:50% mix of $(H_3O^+)[H_2PMo_{12}O_{40}]^-$ and $(H_3O^+)[H_4PMo_{12}O_{40}]^-$ (red). The current density is based on the primary electrodes' area. The plots were reconstructed according to data from [15].*

The position of the mediator's redox wave in between the HER and OER standard potentials enables its oxidation and reduction without parasitic oxygen or hydrogen evolution, thus facilitating complete product separation and eliminating the risk of $H_2/O_2$ mixing, even under partial load conditions.

Nevertheless, a membrane must still be installed when using a soluble redox mediator, to separate the anolyte and catholyte and prevent the reverse redox reaction of the mediator from taking place, unintentionally, at the primary electrode. Without a membrane, the soluble mediator may act as a redox shuttle between the primary and auxiliary electrodes, failing to fulfill the desired purpose of mediating the proton exchange between the HER and OER.

Figure 4 presents three operation modes for decoupled water splitting using a soluble redox mediator in acidic solutions. Conceptually, the same operation modes could be applied in alkaline solutions, with hydroxide ions in place of protons and an anion exchange membrane (AEM) in place of the proton exchange membrane (PEM). However, AEMs are still in the research and development stage and, unlike PEM, they are not commercially available. Furthermore, a suitable soluble redox mediator that is stable in alkaline conditions has yet to be found.

The operation mode presented by Symes and Cronin[15] is represented in the schematic illustration in Figure 4A, wherein the same cell alternates between oxygen generation in one step (step ① in the figure) and hydrogen generation in the other step (step ② in the figure), such that the production of hydrogen is intermittent. For continuous hydrogen production, two units must operate concurrently, one producing hydrogen while the other one produces oxygen, as illustrated in Figure 4B. In this configuration, the two primary electrodes are connected to the power source while the two auxiliary electrodes are electrically connected to each other in short circuit. This doubles the material demand and consequently increases the capital costs. However, decoupled operation may reduce the costs associated with gas separation and membrane degradation, thus potentially offsetting the increase in material expenses. In both the one-cell configuration illustrated in Figure 4A and the two-cell configuration illustrated in Figure 4B, each cell generates hydrogen and oxygen alternately. To operate the system with designated cells that continuously produce only hydrogen or oxygen in each cell, the redox mediator solution can be circulated between the two cells, either in batch mode (complete replacement of the solutions in each cell every cycle), or in a continuous mode as illustrated in Figure 4C.

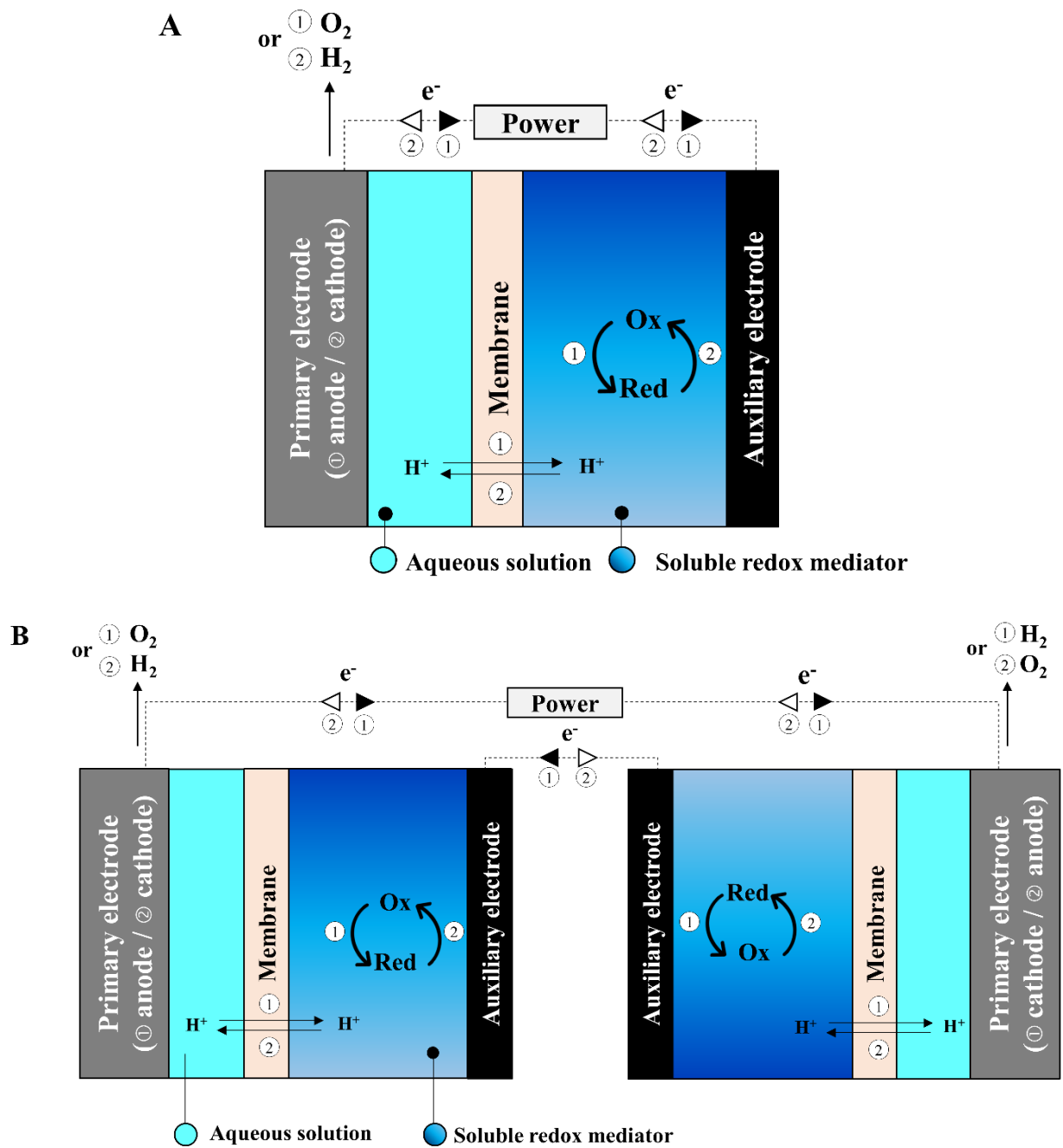

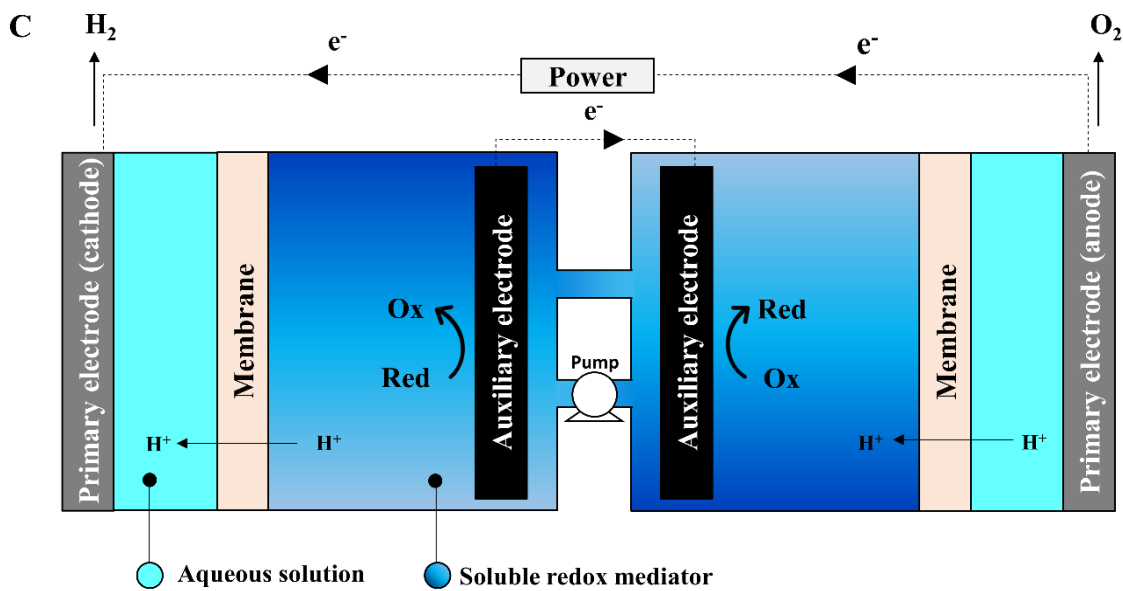

*Figure 4: Schematic illustrations of three operation modes for electrolytic decoupled water splitting in acidic electrolytes using soluble redox mediators that mediate the proton ($H^+$) exchange between the HER and OER. In two-step operation modes the first step and second steps are marked by ① and ②, respectively. (A) One-cell operation mode with alternating oxygen and hydrogen generation steps (① and ②, respectively); (B) Two-cell operation mode with alternating oxygen ① and hydrogen ② cells; (C) Two-cell operation mode with designated hydrogen and oxygen cells and continuous circulation of the redox mediator solution between the two cells.*

Following the ground-breaking work of Symes and Cronin,[15] decoupled water splitting has emerged as a fast growing research field, with more soluble redox mediators being explored. For instance, Rausch *et al.* proposed a quinone derivative, potassium hydroquinone sulfonate (PHS), as a low-cost, bioinspired organic ECBF, with a suitable redox wave between the HER and OER onset potentials.[16] Unfortunately, the cyclability of this quinone derivative was found to be rather limited, losing ~1% of its stored capacity on each cycle. Later, Kirkaldy *et al.* successfully demonstrated stable continuous operation using a different quinone derivative, anthraquinone-2,7-disulfonic acid (AQDS), according to the scheme presented in Figure 4C.[18] AQDS is an organic ECPB, with a relatively low molecular weight (412.3 g/mol, compared to 1825.25 g/mol for PMA), and it has one redox wave within the water stability range, as shown in Figure 5, depicting the CV of AQDS measured with a glassy carbon electrode dipped in an aqueous solution of 25 mM AQDS and 1M $H_2SO_4$. The linear sweep voltammograms (LSV) for a Pt electrode and for a glassy carbon electrode in aqueous electrolytes of $H_2SO_4$ without AQDS are overlaid on the CV curve in Figure 5. Although the AQDS reduction potential is close to the onset potential of hydrogen evolution on a Pt electrode (in 1M $H_2SO_4$), the high overpotential for the HER on a glassy carbon electrode shifts the onset potential of hydrogen evolution well below the AQDS reduction potential and enables the reduction of AQDS without parasitic hydrogen evolution.[18]

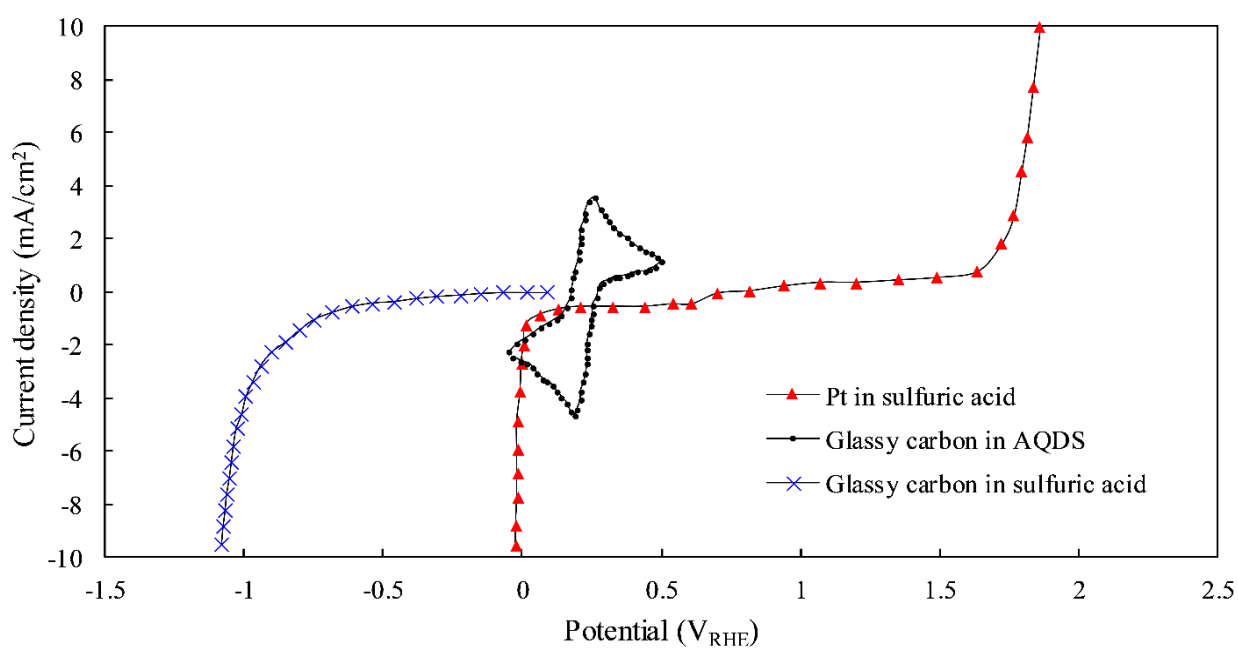

*Figure 5: Cyclic voltammogram (black circles) measured with a glassy carbon working electrode in an aqueous solution of 1M $H_2SO_4$ and 25 mM AQDS, and linear sweep voltammograms measured in aqueous solutions of 1M $H_2SO_4$ with a Pt working electrode (red triangles) and in 0.5M $H_2SO_4$ with a glassy carbon working electrode (blue Xs). Plots reconstructed according to data from* [18,35].

In Kirkaldy's system, each cell comprised of two compartments separated by a catalyst-coated Nafion membrane, wherein each membrane was coated on one side by a suitable catalyst (Ir for the OER and Pt for the HER), while the other side was left uncoated so that the AQDS redox reactions occurred at the surface of bare carbon auxiliary electrodes. The hydrogen and oxygen evolution compartments were filled with an aqueous electrolyte of 1M $H_2SO_4$ (without AQDS), and the soluble redox mediator compartments contained 0.5M AQDS, pre-charged to a state of charge (SOC) of 50%, in addition to 1M $H_2SO_4$. The system was operated galvanostatically at a current density of 250 mA/cm$^2$ while the AQDS electrolyte was continuously circulated between the cells to allow stable operation around 50% SOC in both cells. At this current density, the hydrogen and oxygen cell voltages were about 0.5 and 1.5 V, respectively, corresponding to an overall voltage efficiency of 1.48V / (0.5+1.5)V = 74%$_{HHV}$.

Another continuous-flow system was demonstrated by Chisholm *et al.*, using silicotungstic acid (STA, $H_4(SiW_{12}O_{40})$) as a soluble redox mediator.[21] This redox mediator was first explored by Rausch *et al.*[22] in an electrochemical – chemical decoupled water splitting scheme, and its properties are discussed in detail previously. Similarly to AQDS, STA is able to perform redox chemistry at a carbon-based auxiliary electrode without concurrent hydrogen generation, due to the high overpotential for hydrogen evolution at this electrode. The system demonstrated by Chisholm *et al.* comprised of two PEM cells connected in series, as in the general scheme presented in Figure 4C, with $IrO_2$ OER catalyst, Pt HER catalyst and carbon paper auxiliary electrodes. Under optimized conditions, the system reached steady state with an oxygen cell voltage of 2.16 V at a current density of 500 mA/cm$^2$, and a Faradaic efficiency of 99% for the STA reduction reaction. This corresponds to what the authors termed the *"decoupling efficiency"*, *i.e.*, the extent to which the HER and OER can truly be decoupled using the redox mediator. The hydrogen cell was operated at the same current as the oxygen cell. Unfortunately, the voltage applied to this cell was not reported, so the overall energy conversion efficiency of this system cannot be calculated.

Chisholm *et al.* validated the robustness of their decoupled system compared to an equivalent coupled system wherein the OER and HER occur concurrently in the same cell. First, the authors showed that at lower current densities of 25 and 50 mA/cm$^2$, the extent of hydrogen cross-permeation through the Nafion membrane was significantly lower in the decoupled system (0.31% $H_2$ in the $O_2$ stream) than in the equivalent coupled one (1.47-1.89% $H_2$ in the $O_2$ stream). Second, the stability of the membrane in the decoupled system was demonstrated by comparing the fluoride concentration in the electrolyte solution (as a measure of the extent of membrane degradation) between the decoupled and coupled systems. In conventional electrolyzers, hydrogen cross-permeation through the membrane can lead to the formation of reactive oxygen species (*e.g.*, $H_2O_2$ and free radicals) that attack the Nafion membrane, leading to its eventual degradation and loss of proton conductivity.[11] Here, the authors showed that this can be avoided in a decoupled system wherein the extent of hydrogen generation in the oxygen cell was negligible, measuring a fluoride concentration of 0.018 mg fluoride/l after 213 h of operation in the decoupled system, compared to 0.5 mg fluoride/l in its coupled counterpart. Finally, the authors also showed an improved durability of the membrane in the decoupled system in the presence of impurities in mineralized water.[21] This improvement was attributed to the acidic nature of the STA mediator, which provides a route for ion exchange with impurities in the electrolyte.

The reported soluble redox mediators that enable pH buffering are stable in acidic conditions only. Proton-independent redox mediators, which undergo redox chemistry without reversible take-up of protons, were proposed by Li *et al.* using the (ferrocenylmethyl)trimethylammonium chloride (FcNCl) couple,[36] and by Goodwin and Walsh using the potassium ferri/ferrocyanide $K_3Fe(CN)_6/K_4Fe(CN)_6$ couple.[17] On the one hand, these proton-independent redox mediators are stable over a wider pH range than the abovementioned ECPBs (PMA, PHS, ADQS and STA), and enable water splitting at near-neutral conditions. On the other hand, their lack of pH buffering ability results in a pH gradient and an overall pH swing throughout operation. Further research is required to identify pH-buffering soluble redox mediators that are stable in a wider pH range, and specifically in alkaline solutions.

**Electrolytic water splitting with solid auxiliary redox electrodes**

As their name suggests, soluble redox mediators are dissolved in the aqueous electrolyte and their redox reactions are carried out at inert auxiliary electrodes (*e.g.*, carbon). To keep the redox mediator from forming a redox shuttle between the primary and auxiliary electrodes, an ion-exchange membrane (*e.g.*, Nafion for acidic electrolytes) separates the anolyte from the catholyte within the electrolytic cell (both in the hydrogen and oxygen cells). By replacing the soluble redox mediator with a solid auxiliary redox electrode that mediates the ion exchange (protons in acidic electrolytes and hydroxide ions in alkaline electrolytes) between the HER and OER, the need for a membrane or separator is eliminated.

This path was first proposed by Rothschild *et al.* in a patent application that was filed in 2014,[23] followed by a scientific article by Landman *et al.* that was published in 2017.[25] A similar approach was reported by Chen *et al.* in 2016.[24] In both studies, nickel hydroxide / oxyhydroxide ($Ni(OH)_2$ / $NiOOH$) electrodes were used as hydroxide-mediating auxiliary redox electrodes in alkaline solutions. Nickel hydroxide is commonly used as the active material in positive electrodes (*i.e.*, cathodes) of rechargeable (secondary) alkaline batteries such as Ni-Cd, Ni-MH and Ni-H$_2$ batteries,[37–39] where it reversibly oxidizes to nickel oxyhydroxide (NiOOH):

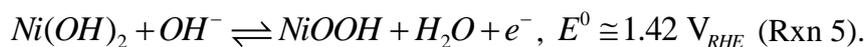

$$Ni(OH)_2 + OH^- \rightleftharpoons NiOOH + H_2O + e^-, \quad E^0 \cong 1.42 \text{ V}_{RHE} \quad \text{(Rxn 5)}.$$

In contrast to the soluble ECPBs described in the previous section, which are stable in acidic electrolytes that require rare Earth catalysts such as Pt and Ir or Ru for the HER and OER, respectively, nickel (oxy)hydroxide is stable in alkaline solutions wherein non-precious metal catalysts such as Ni-Fe layered double hydroxide (LDH) and Raney nickel can be used as OER and HER catalysts, respectively.[40,41] Additionally, since nickel (oxy)hydroxide undergoes redox cycling in the solid state, no membrane or separator is required, and a single electrolyte solution fills the cell. Some critical disadvantages of alkaline electrolyzers, such as hydrogen cross-permeation through the separator at partial loads or high pressures and sensitivity to impurities at the electrolyte, stem from the coupled operation and the use of a separator.[2,3,12] Decoupled membraneless water splitting systems with hydroxide-mediating auxiliary redox electrodes could potentially overcome these problems, paving a new path towards high-pressure hydrogen production and partial load operation.

For hydroxide-mediating redox electrodes in alkaline solutions, guideline (i) for the selection of suitable redox mediators (see list in pp. 10-11) is expanded to include mechanical durability, in addition to chemical stability. Guideline (vii) requires high capacity per unit volume and/or weight of the redox electrode, and guideline (viii) requires that the only counter ion in the mediator's redox reactions be hydroxide ($OH^-$). Nickel (oxy)hydroxide electrodes comply with all of the selection criteria for hydroxide-mediating redox electrodes, being stable in aqueous alkaline solutions with the only counter ion for their redox reactions being a hydroxide ion, and having excellent cycling durability, enabling tens of thousands of charge-discharge cycles as demonstrated in Ni-$H_2$ batteries.[37,42,43]

Nickel (oxy)hydroxides are generally classified into four phases, as presented in Figure 6: Two hydroxides, α-Ni(OH)$_2$ and β-Ni(OH)$_2$, and two oxyhydroxides, β-NiOOH and γ-NiOOH.[42] β-Ni(OH)$_2$ is the discharged phase that is stable in alkaline solutions, and it can be reversibly charged to the β-NiOOH phase without significant changes in the crystal volume. However, overcharging can lead to formation of the γ-NiOOH phase, which has a significantly larger interplanar spacing, and therefore causes swelling and fast deterioration of electrode performance upon cycling. For these reasons, nickel (oxy)hydroxide electrodes are best cycled in the β-Ni(OH)$_2$ / β-NiOOH cycle by avoiding overcharging.[37,44]

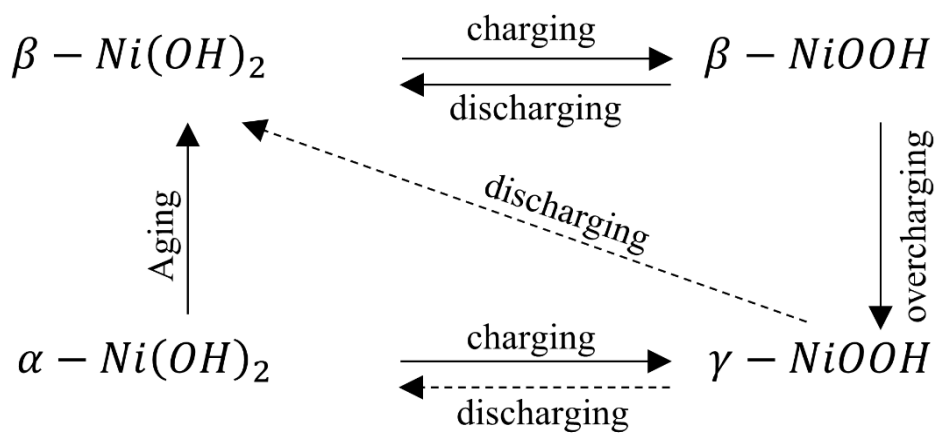

*Figure 6: The Bode scheme of the phase transformations of nickel (oxy)hydroxide in alkaline solution.*[42,44]

The standard potential of the Ni(OH)$_2$ / NiOOH redox couple is 1.42 V$_{RHE}$,[45] though the reversible potential of Ni(OH)$_2$-based electrodes may differ from this value by several tens of mV depending on their chemical composition, microstructure and phase.[46,47] Figure 7 shows a cyclic voltammogram of the β-Ni(OH)$_2$ electrode used by Landman *et al.*[25] It can be seen that although the redox potential of this electrode (~1.35 V$_{RHE}$) is higher than the reversible OER potential (1.23 V$_{RHE}$), its redox wave precedes the onset potential of oxygen evolution by approximately 150 mV, owing to the large OER overpotential. This kinetic suppression of the OER enables Ni(OH)$_2$ oxidation without parasitic oxygen evolution under the appropriate operation conditions.

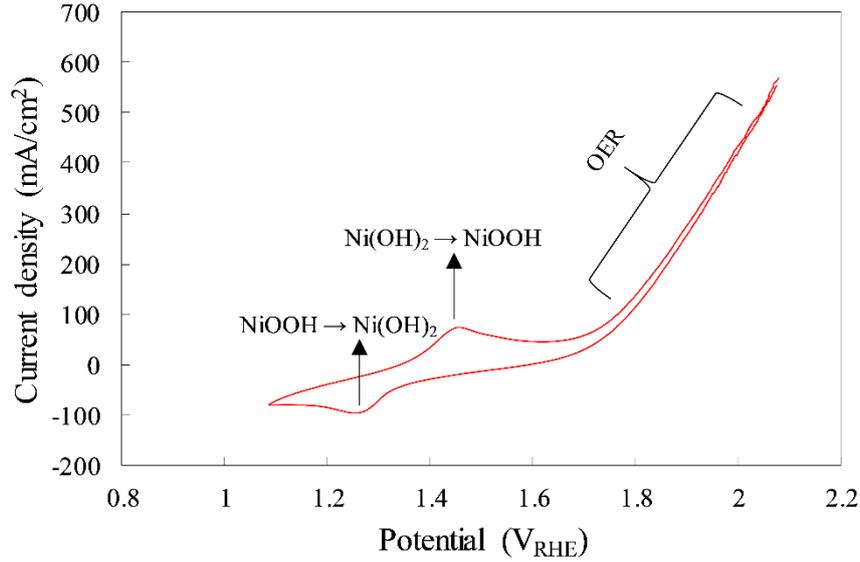

*Figure 7: Cyclic voltammogram of a β-Ni(OH)₂ electrode in 1M NaOH aqueous electrolyte (at ambient temperature). Plot reconstructed according to data from* [25].

In analogy to the soluble ECPBs described in the previous section, which operate in acidic solutions, the nickel (oxy)hydroxide electrode serves as an electron-coupled hydroxide buffer (ECHB) in alkaline solutions, where it mediates the exchange of hydroxide (OH⁻) ions between the primary electrodes. During the hydrogen generation step, hydrogen is produced at the primary cathode while the Ni(OH)₂ auxiliary electrode is oxidized to NiOOH:

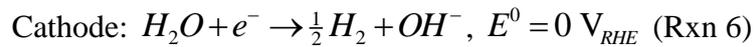

Cathode: $H_2O + e^- \rightarrow \tfrac{1}{2} H_2 + OH^-$, $E^0 = 0$ V$_{RHE}$ (Rxn 6)

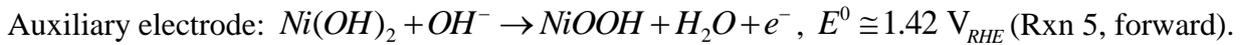

Auxiliary electrode: $Ni(OH)_2 + OH^- \rightarrow NiOOH + H_2O + e^-$, $E^0 \cong 1.42$ V$_{RHE}$ (Rxn 5, forward).

Here, the hydroxide ions produced by the HER travel through the alkaline (1M KOH or NaOH)[24,25] aqueous electrolyte, without a membrane or separator, to the Ni(OH)₂ auxiliary electrode, where they are taken up by the Ni(OH)₂ oxidation reaction to NiOOH. During the oxygen generation step, oxygen is produced at the primary anode while the oxidized NiOOH auxiliary electrode is reduced back to its initial state, Ni(OH)₂:

Anode: $OH^- \rightarrow \tfrac{1}{4}O_2 + \tfrac{1}{2}H_2O + e^-$, $E^0 = 1.23$ $V_{RHE}$ (Rxn 7)

Auxiliary electrode: $NiOOH + H_2O + e^- \rightarrow Ni(OH)_2 + OH^-$, $E^0 \cong 1.42$ $V_{RHE}$ (Rxn 5, backward).

The hydroxide ions travel through the electrolyte from the NiOOH auxiliary electrode to the primary anode where they are taken up by the OER (Rxn 7). Although the standard potentials of Rxns 5 and 7 suggest that this step may proceed spontaneously with a negative cell voltage of -0.19 V (as in galvanic cells), the slow kinetics of the OER with respect to the fast kinetics of the more facile NiOOH reduction reaction (Rxn 5, backward) give rise to overpotentials that inhibit this step from occurring spontaneously at near ambient temperatures. Therefore, this step requires the application of a small external bias to proceed. Later we will see that this scenario changes at elevated temperatures.

Figure 8 presents three operation modes for electrolytic decoupled water splitting in alkaline solutions using hydroxide-mediating auxiliary redox electrodes. These operation modes could also be applied to acidic solutions, with protons ($H^+$) in the electrolyte in place of the hydroxide ions ($OH^-$) depicted. Figure 8A illustrates a one-cell operation mode in which hydrogen and oxygen are produced sequentially in the same cell, as reported by Chen *et al.*,[24] whereas Figures 8B and 8C illustrate two-cell operation modes in which hydrogen and oxygen are produced concurrently at two different cells, as reported by Landman *et al.*[25] In Figures 8B and 8C, two cells are connected in series, such that in the first cycle one cell produces hydrogen at the primary cathode while the auxiliary redox electrode is oxidized, whereas the second cell produces oxygen at the primary anode while the auxiliary redox electrode is reduced. At the end of the first cycle, the auxiliary electrodes in the hydrogen and oxygen cells are oxidized and reduced to their full operational capacity, respectively. At this point, two options are available to continue the operation: (i) Reversing the current polarity and alternating the cell functions such that the cell that produced hydrogen in the first cycle produces oxygen in the second cycle, and vice versa (Figure 8B); or (ii) Swapping the auxiliary electrodes from one cell to another (Figure 8C).

In Figure 8A and 8B, the primary electrodes, *i.e.*, the anode and cathode, alternate their function from cycle to cycle, from oxygen generation to hydrogen generation and vice versa. This can be achieved using a couple of primary electrodes within each cell, one functionalized with OER catalyst and the other one with HER catalyst, and a switch that selects the right electrode for each function. Alternatively, bifunctional catalysts that catalyze both the OER and HER can be used.[48] In this case, a single primary electrode in each cell suffices, as illustrated in Figures 8A-C.

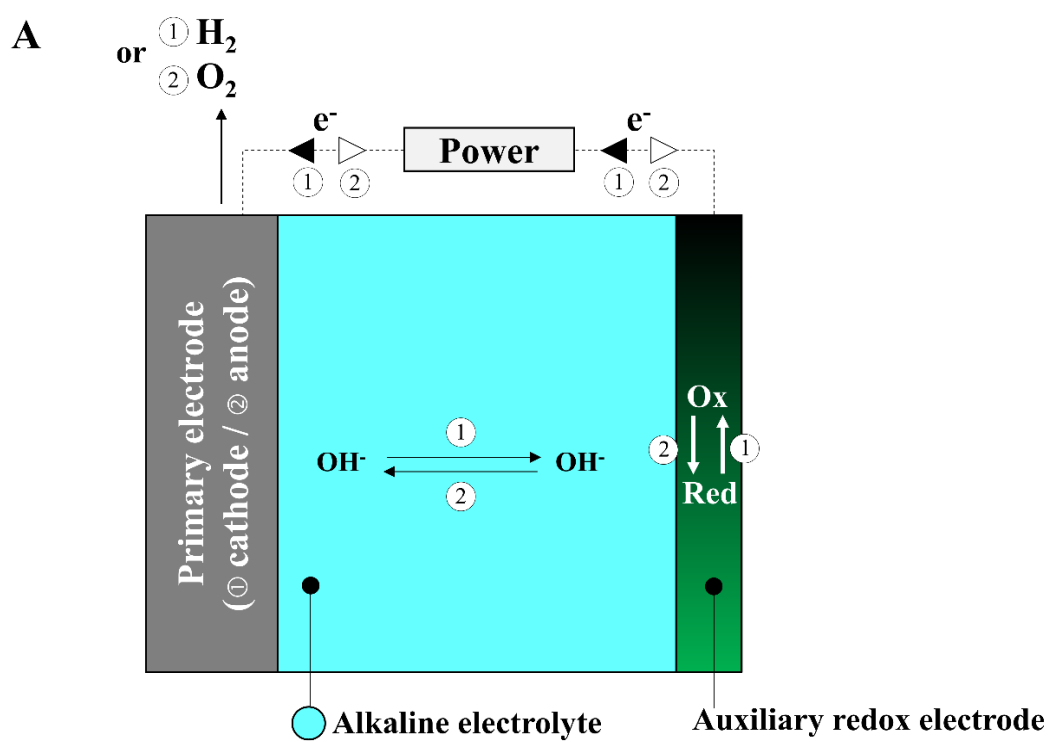

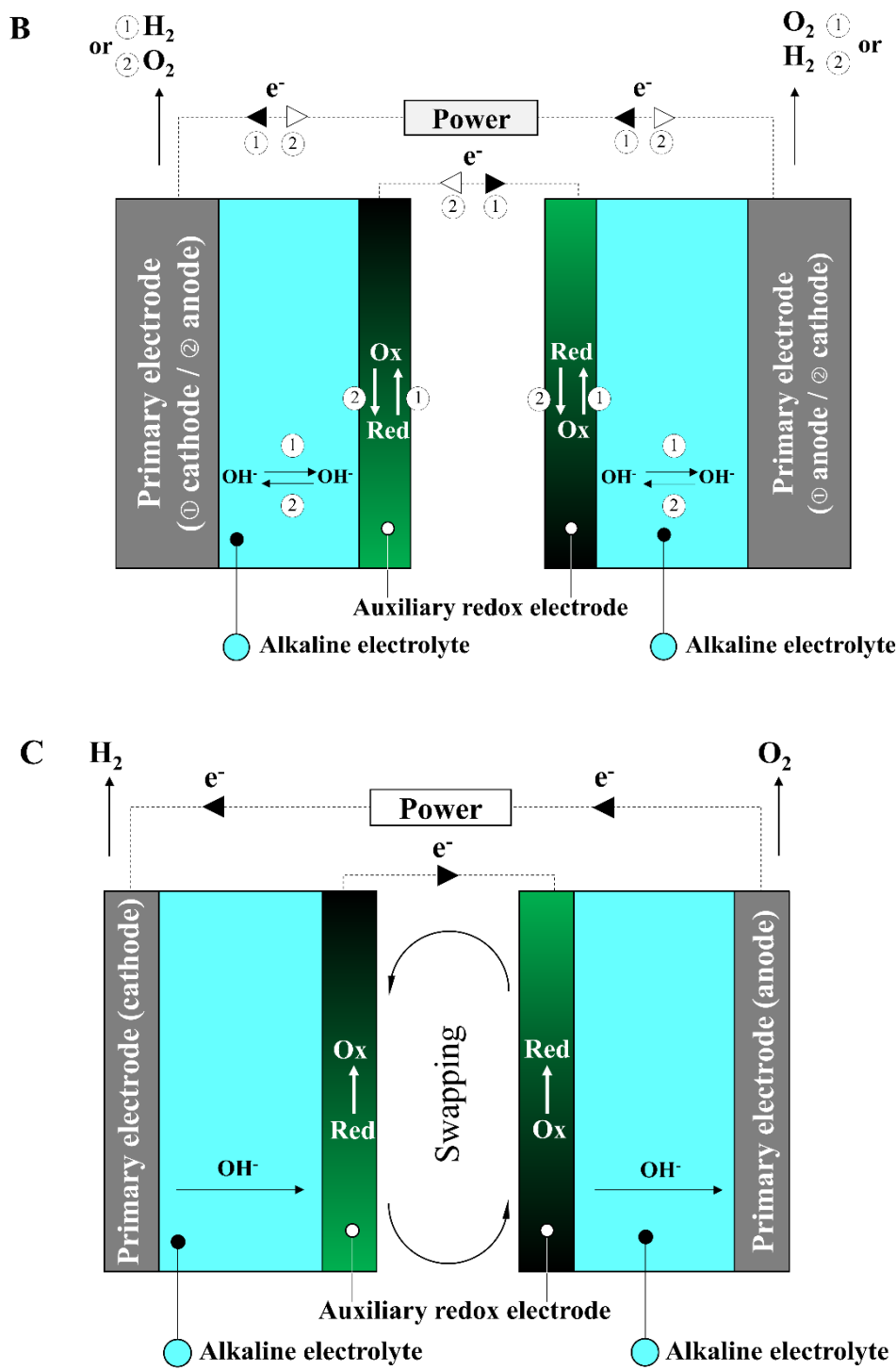

*Figure 8: Schematic illustrations of three operation modes for decoupled water splitting in alkaline aqueous electrolytes using auxiliary redox electrodes that mediate the hydroxide ($OH^-$) exchange between the HER and OER. In two-step operation modes the first and second steps are marked by ① and ②, respectively. (A) One-cell operation mode with sequential hydrogen generation ① and oxygen generation ② steps; (B) Two-cell operation mode with cycling by current polarity reversal that alternates the hydrogen and oxygen cells; (C) Two-cell operation mode with designated hydrogen and oxygen cells and cycling by auxiliary electrode swapping.*

As opposed to soluble redox mediators, which may be circulated between the hydrogen and oxygen cells for continuous operation as illustrated in Figure 4C, the application of solid auxiliary redox electrodes only allows for batch-mode or swing-mode operation as illustrated in Figures 8A and 8B, respectively. In a completely electrolytic scheme, the current polarity could be reversed, accompanied by periodic purging of the cells between cycles to prevent $H_2/O_2$ mixing. However, in applications where current polarity reversal is not possible such as photoelectrochemical water, the auxiliary redox electrodes must be swapped, as illustrated in Figure 8C, a task that increases the system's complexity and cost.[25]

Landman *et al.*[25] demonstrated forty decoupled water splitting alternating cycles using battery-grade nickel (oxy)hydroxide redox electrodes according to the operation scheme presented in Figure 8B. Figure 9A presents the applied voltage as a function of time during these cycles. An average voltage of 2.12 V was achieved at a current density of 5 mA cm$^{-2}$ using Ni foil primary electrodes in 1M NaOH aqueous electrolyte, corresponding to a voltage efficiency of 1.48V / 2.12V = 70%$_{HHV}$. The applied voltage in the equivalent coupled system with the same primary electrodes was 2V, such that the decoupling loss was only 1 - 2V / 2.12V = 5.7%.

The charge that was transferred in each cycle decreased from one cycle to another, as shown in Figure 9B. This drift resulted from incomplete charging of the auxiliary redox electrode that accumulated from one cycle to another in this particular mode of operation. A small percentage (0.33% in this case) of the charge $Q$ that was passed during charging the Ni(OH)$_2$ electrode was lost, in each cycle, due to parasitic reactions. Hence, only $0.9967Q$ could be extracted in the next cycle, and so forth on subsequent cycles. After *n* cycles, the charge that could be extracted from the auxiliary redox electrode during discharging reduced to $(0.9967^n)Q$, reaching 88% of the initial charge ($Q$) after forty cycles. It is possible to overcome this drift and achieve perfectly stable cycles in the two-cell operation mode by operating the hydrogen cell a bit longer than the oxygen cell. As a result, an additional charge is passed to the charging auxiliary electrode, compensating for the lost charge due to the parasitic reactions.

Using this method, Landman *et al.* demonstrated five cycles with no drift, as shown in the inset of Figure 9B.

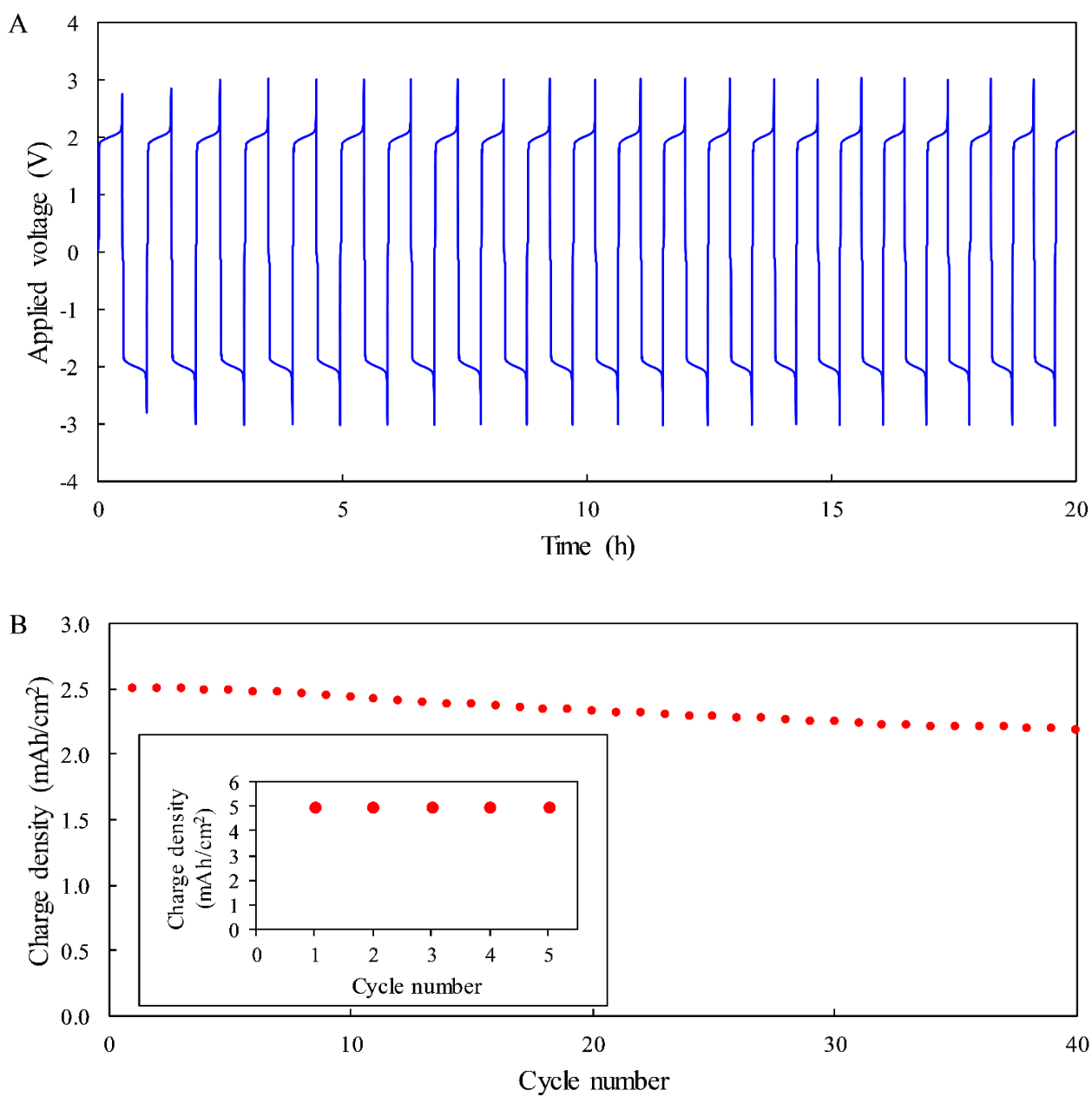

*Figure 9: Decoupled water splitting cycles with battery-grade nickel (oxy)hydroxide auxiliary redox electrodes in 1M NaOH aqueous electrolyte at ambient conditions, with Ni foil primary electrodes (both anode and cathode). (A) The applied voltage during forty alternating cycles carried out by current polarity reversal at a current density of 5 mA/cm$^2$ in two cells connected in series, according to the scheme in Figure 8B. (B) The charge that was passed in each of the forty cycles as a function of cycle number. The inset shows five cycles with no drift, achieved by operating the hydrogen cell slightly longer than the oxygen cell, as explained in the text. Plots reconstructed according to data from [25].*

Besides nickel (oxy)hydroxide, manganese dioxide ($MnO_2$) was proposed by Choi *et al.* as a candidate redox electrode that mediates the hydroxide ($OH^-$) exchange between the HER and OER in alkaline solutions.[26] Similarly to $Ni(OH)_2$, $MnO_2$ undergoes a reversible oxidation to manganese oxyhydroxide (MnOOH). Its main advantages, compared to $Ni(OH)_2$, are its lower redox potential (0.977 $V_{RHE}$), reducing the risk for parasitic oxygen evolution during charging, and higher theoretical capacity (308 mAh/g, compared to 289 mAh/g for $Ni(OH)_2$). However, its main disadvantage is the possible formation of the irreversible $Mn_3O_4$ phase upon discharge, which limits the number of redox cycles that this material can undergo.[49]

Solid redox electrodes in acidic solutions have also been reported. Ma *et al.* and Wang *et al.* proposed the electroactive conducting polymers polytriphenylamine (PTPAn)[27] and polyaniline[29] as proton-mediating redox electrodes in 0.5M $H_2SO_4$ aqueous solution. However, both materials are not ECPBs since their redox reactions do not involve protons. Pyrene-4,5,9,10-tetraone (PTO), an organic substance, was also suggested by Ma *et al.* (in 0.5M $H_2SO_4$).[28] PTO has a higher capacity (156 mAh/g, compared to 77 mAh/g for PTPAn), and is an ECPB, reversibly taking up protons and electrons upon redox cycling.

**Electrochemical – chemical cycles for decoupled water splitting**

The electrolytic decoupling schemes described in the previous sections present some prospective advantages such as simple cell construction and maintenance, especially for membraneless operation schemes using solid auxiliary redox electrodes, with potential to operate at high pressures and intermittent power supply. However, the voltage of the overall decoupled process, $V_{decoupled}$, is greater than the voltage requirement for the equivalent coupled water splitting process, $V_{coupled}$ (see Eq 1). The difference between $V_{decoupled}$ and $V_{coupled}$ stems primarily from the polarization losses of the redox mediator/electrode, *i.e.*, its oxidation and reduction overpotentials ($\eta_{ox} + \eta_{red}$). Therefore, in any scheme that is based on electrochemical oxidation and reduction of a redox mediator/electrode, the energy conversion efficiency of a decoupled water splitting system is *lower* than that of the equivalent coupled system with the same primary electrodes, electrolyte, cell structure, and under the same operation conditions.

This drawback can be lifted in an electrochemical - chemical decoupling scheme, wherein one of the redox reactions (either oxidation or reduction) of the redox mediator/electrode is carried out spontaneously, *i.e.*, chemically (instead of electrochemically), without consuming electrical power. For example, the redox mediator/electrode may undergo electrochemical oxidation during the hydrogen generation step (as in the electrolytic decoupling schemes described previously), followed by a spontaneous chemical reduction that is catalyzed by some control parameter such as temperature or catalyst. Hence, not only is the reduction overpotential ($\eta_{red}$) of the redox mediator/electrode taken out of the equation, but the OER overpotential is replaced by the overpotential of the oxidation reaction of the redox mediator/electrode, which may be lower than the OER overpotential. This may lead to a lower voltage (during the electrochemical step) than the voltage applied in conventional water electrolysis, thereby offering higher energy conversion efficiency in addition to the other operational advantages of decoupled water splitting that were discussed above. Consequently, electrochemical - chemical decoupled water splitting offers potential advantages over conventional water electrolysis in both operational and energy saving aspects.

**Electrochemical - chemical water splitting with soluble redox couples**

Rausch *et al.* reported an electrochemical – catalyst-activated chemical (ECAC) decoupling strategy using silicotungstic acid (STA, $H_4(SiW_{12}O_{40})$) as an ECPB redox couple in an acidic electrolyte (1M $H_3PO_4$ in the anodic compartment, and 0.5M $H_4(SiW_{12}O_{40})$, pH 0.5, in the cathodic compartment).

Figure 10 shows CV of 0.5M STA aqueous solution measured with a glassy carbon working electrode under Ar purge, displaying two redox waves centered around +0.01 (wave I, $H_4(SiW_{12}O_{40}) + H^+ + e^- \rightleftharpoons H_5(SiW_{12}O_{40})$, Rxn 8) and -0.22 $V_{RHE}$ (wave II, $H_5(SiW_{12}O_{40}) + H^+ + e^- \rightleftharpoons H_6(SiW_{12}O_{40})$, Rxn 9). Overlaid on the CV scan are cathodic LSVs measured with a glassy carbon working electrode or a Pt disc working electrode in 1M $H_3PO_4$ aqueous solution. Without STA, the HER onset at the glassy carbon electrode occurs at approximately -0.6 $V_{RHE}$. Since both the redox waves of STA are anodic to -0.6 $V_{RHE}$, it can be reduced from $H_4(SiW_{12}O_{40})$ to $H_6(SiW_{12}O_{40})$ at a glassy carbon electrode without concurrent hydrogen evolution.

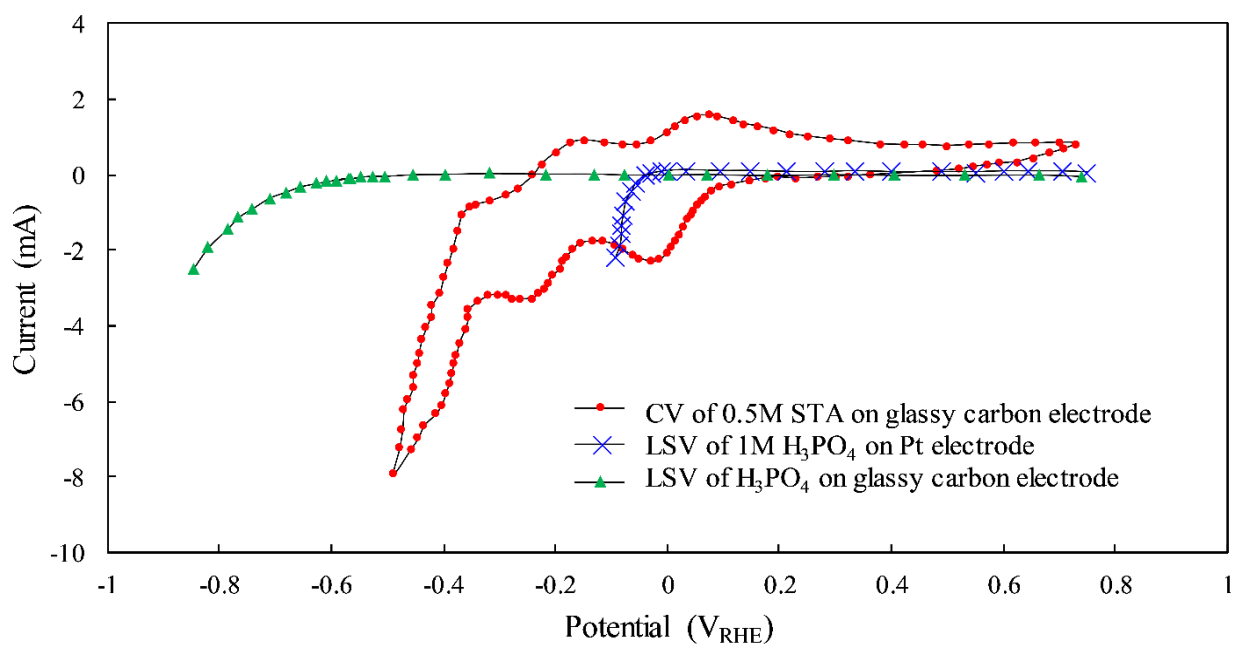

*Figure 10: Cyclic voltammogram measured with a glassy carbon working electrode in 0.5M STA aqueous solution at ambient temperature (red circles), alongside linear sweep voltammograms measured with a glassy carbon working electrode (green triangles) and a Pt disc working electrode (blue Xs) in 1M $H_3PO_4$ aqueous electrolyte. All plots reconstructed according to data from [22].*

In a completely electrolytic decoupling scheme, $H_4(SiW_{12}O_{40})$ would be electrochemically reduced at a glassy carbon electrode during the oxygen generation step, as its redox waves are anodic to the onset potential for hydrogen evolution at this electrode, and then oxidized electrochemically during the hydrogen generation step. This mode of operation was discussed previously. However, in the presence of Pt catalyst, the two-electron reduced form of STA, $H_6(SiW_{12}O_{40})$, can be spontaneously oxidized to the one-electron reduced form, $H_5(SiW_{12}O_{40})$, with concurrent hydrogen evolution according to the chemical reaction:

$$H_6(SiW_{12}O_{40}) \xrightarrow{Pt} H_5(SiW_{12}O_{40}) + \tfrac{1}{2}H_2 \quad (\text{Rxn 10}).$$

This reaction occurs spontaneously, *i.e.*, without applying external bias, since its free energy change is negative (exergonic reaction), as can be seen by the position of the respective redox wave (wave II) in Figure 10, which is cathodic to the onset potential of hydrogen evolution at the Pt electrode. This suggests that while the spontaneous chemical re-oxidation of the two-electron reduced form, $H_6(SiW_{12}O_{40})$, is kinetically inhibited in the presence of glassy carbon, it becomes facile in the presence of a Pt catalyst.

Thus, the STA redox couple can be reduced at a glassy carbon electrode to yield the fully (two-electron) reduced form, $H_6(SiW_{12}O_{40})$, without competing hydrogen evolution, which can then be stored without spontaneous hydrogen evolution. Finally, in the presence of a Pt catalyst, the fully reduced STA reduces water to form hydrogen, while being itself oxidized to $H_5(SiW_{12}O_{40})$ in a spontaneous chemical reaction. This enables an electrochemical – chemical water splitting cycle, in which the first step is electrochemical, wherein oxygen is generated at the anode while the STA redox couple is reduced at a glassy carbon electrode:

$$\text{Anode: } H_2O \rightarrow \tfrac{1}{2}O_2 + 2H^+ + 2e^-, \; E^0 = 1.23 \text{ V}_{RHE} \quad (\text{Rxn 1})$$

$$\text{Glassy carbon electrode: } H_4(SiW_{12}O_{40}) + H^+ + e^- \rightarrow H_5(SiW_{12}O_{40}), \; E^0 \cong 0.01 \text{ V}_{RHE} \quad (\text{Rxn 8})$$

$$H_5(SiW_{12}O_{40}) + H^+ + e^- \rightarrow H_6(SiW_{12}O_{40}), \; E^0 \cong -0.22 \text{ V}_{RHE} \quad (\text{Rxn 9}).$$

The reduction to the fully reduced form, $H_6(SiW_{12}O_{40})$, occurs (on the glassy carbon electrode) at potentials positive of -0.6 $V_{RHE}$. At these potentials no hydrogen evolves at the glassy carbon electrode (see Figure 10). Therefore, only oxygen is generated in this step.

The second step is then carried out chemically, *i.e.*, without applying external bias, by exposing the fully (two electron) reduced STA to a Pt catalyst, inducing its spontaneous re-oxidation to the partially (one electron) reduced STA with concomitant hydrogen evolution:

$$H_6(SiW_{12}O_{40}) \xrightarrow{Pt} H_5(SiW_{12}O_{40}) + \tfrac{1}{2}H_2 \quad \text{(Rxn 10)}.$$

Figure 11 illustrates the general operation scheme proposed by Rausch *et al.* using a soluble redox couple.

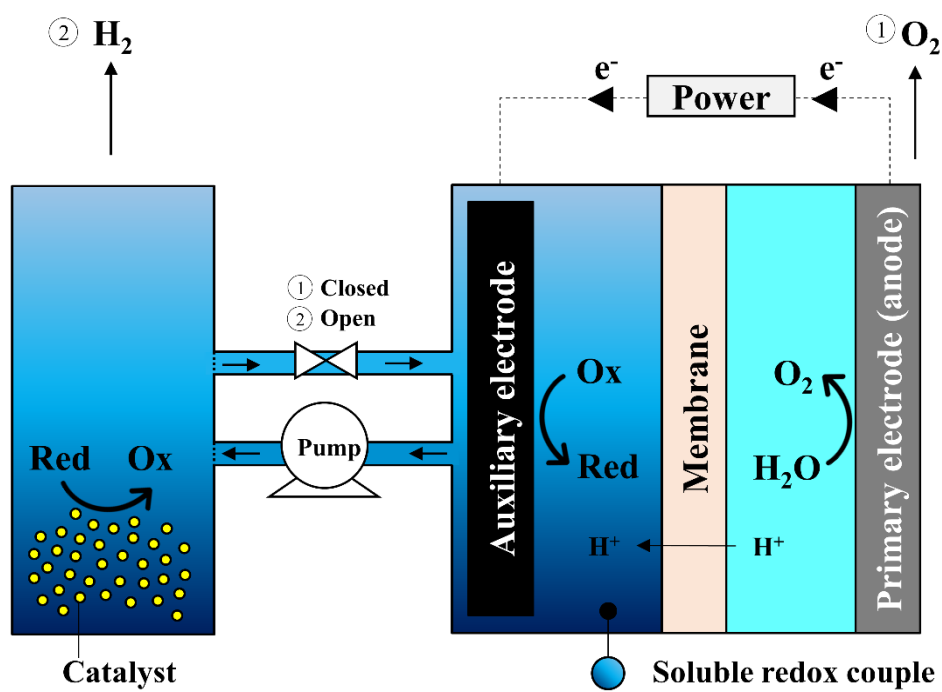

*Figure 11: Illustration of an electrochemical – catalyst-activated chemical decoupled water splitting scheme with a soluble redox couple, following the system proposed by Rausch et al.[22] According to this scheme, water splitting is carried out in two steps. First, the soluble redox couple is reduced at an auxiliary electrode while water is oxidized at the primary electrode (anode). Then, the catholyte containing the reduced redox couple is circulated to another cell containing a suitable catalyst. Once in contact with the catalyst, the reduced redox couple is chemically and spontaneously oxidized to its initial state (regenerated), releasing hydrogen. The first step and second steps are marked by ① and ②, respectively.*

Because the redox couple's regeneration reaction is chemical, rather than electrochemical, there is no need for current collection at the Pt catalyst. This allows the Pt catalyst to be dispersed as particles in the solution, as illustrated in Figure 11, thereby significantly increasing its active surface area. Indeed, Rausch *et al*. reported that the initial hydrogen generation rate per mg of Pt used exceeded that of PEM electrolysis, reaching initial rates as high as 2861 mmol $H_2$ $h^{-1}$ $mg^{-1}$ (compared to 50-100 mmol $H_2$ $h^{-1}$ $mg^{-1}$ in PEM electrolysis).[22] As shown in Figure 12, most of the hydrogen was generated during the first minute of regeneration. Since hydrogen is the desired product, the duration of any other process, including electrolyte circulation and oxygen generation, should be minimized. Thus, if 0.01 mol of $H_6(SiW_{12}O_{40})$ is chemically oxidized to $H_5(SiW_{12}O_{40})$ in 60 s during the hydrogen generation step, the following oxygen generation step should reduce it back to $H_6(SiW_{12}O_{40})$ in less than 60 s, corresponding to an anodic current of more than 16 A.

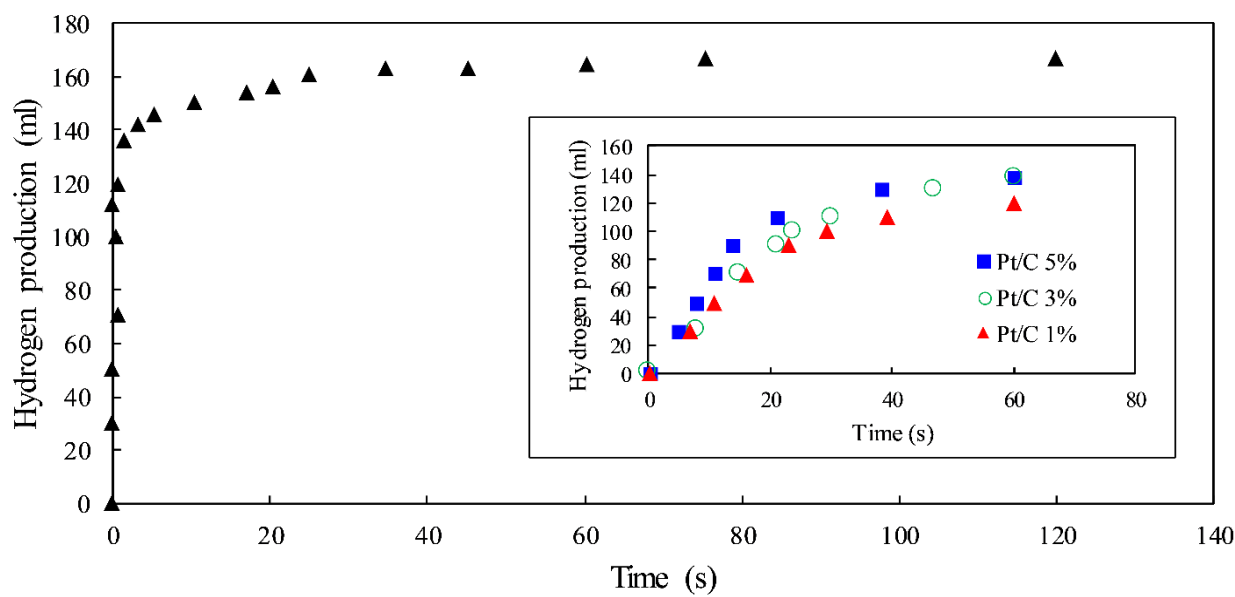

*Figure 12: The hydrogen production from a 20 ml sample of 0.5M STA aqueous solution during the regeneration (re-oxidation) of the reduced $H_6(SiW_{12}O_{40})$ redox couple back to its one-electron oxidized form $H_5(SiW_{12}O_{40})$ when in contact with 50 mg of Pt/C (5 wt.%) catalyst. The inset shows the initial hydrogen production for various Pt loadings (1, 3, and 5 wt.% on C). Plots reconstructed according to data from [22].*

In terms of energy conversion efficiency, the comparison of the ECAC system reported by Rausch *et al.* to the equivalent coupled system requires a careful definition of the primary electrodes. In their decoupled system, Rauch *et al.* used a glassy carbon auxiliary electrode to reduce the STA redox couple, but they also used Pt catalyst for its chemical re-oxidation that evolved hydrogen. To achieve a current density of 50 mA cm$^{-2}$, the voltage required to oxidize water at a Pt anode and to reduce the STA redox couple at a glassy carbon electrode was 2.37 V, corresponding to a voltage efficiency of 1.48V / 2.37V = 62%$_{HHV}$. At the same current density, the voltage required to split water using two Pt electrodes in an electrolytic cell was 2.21 V, corresponding to a voltage efficiency of 67%$_{HHV}$, 5% higher than the decoupled system. Therefore, the gains in system flexibility and hydrogen production rate per mg Pt should be weighed against the loss in efficiency for eventual cost optimization.

Following the pioneering work of Rausch *et al.*,[22] Chen *et al.* reported a similar approach using the polyoxoanion $[P_2W_{18}O_{62}]^{6-}$ as a soluble ECPB redox couple.[19] As shown in Figure 13, $[P_2W_{18}O_{62}]^{6-}$ has several redox waves, corresponding to the passage of 18 electrons per cluster. Four of these redox waves lie cathodically to the reversible HER, corresponding to the passage of 16 electrons per cluster. $[P_2W_{18}O_{62}]^{6-}$ may therefore be used as a soluble redox couple in an electrochemical - chemical cycle using Pt catalyst for re-oxidation and hydrogen generation, similarly to the approach presented by Rausch *et al.* using the STA redox couple.[22] Interestingly, Chen *et al.* found that spontaneous re-oxidation and hydrogen evolution can also be achieved without Pt catalyst, by diluting the solution and raising its pH.[19] At the solubility limit of $[P_2W_{18}O_{62}]^{6-}$, 1.9M in aqueous solution,[19] and delivering 16 electrons per cluster, this redox couple has a hydrogen storage capacity of 30.4 g H$_2$ l$^{-1}$. By comparison, a 0.5M aqueous solution of STA delivering one electron per cluster below the HER standard potential has a storage capacity of only 0.5 g H$_2$ l$^{-1}$.

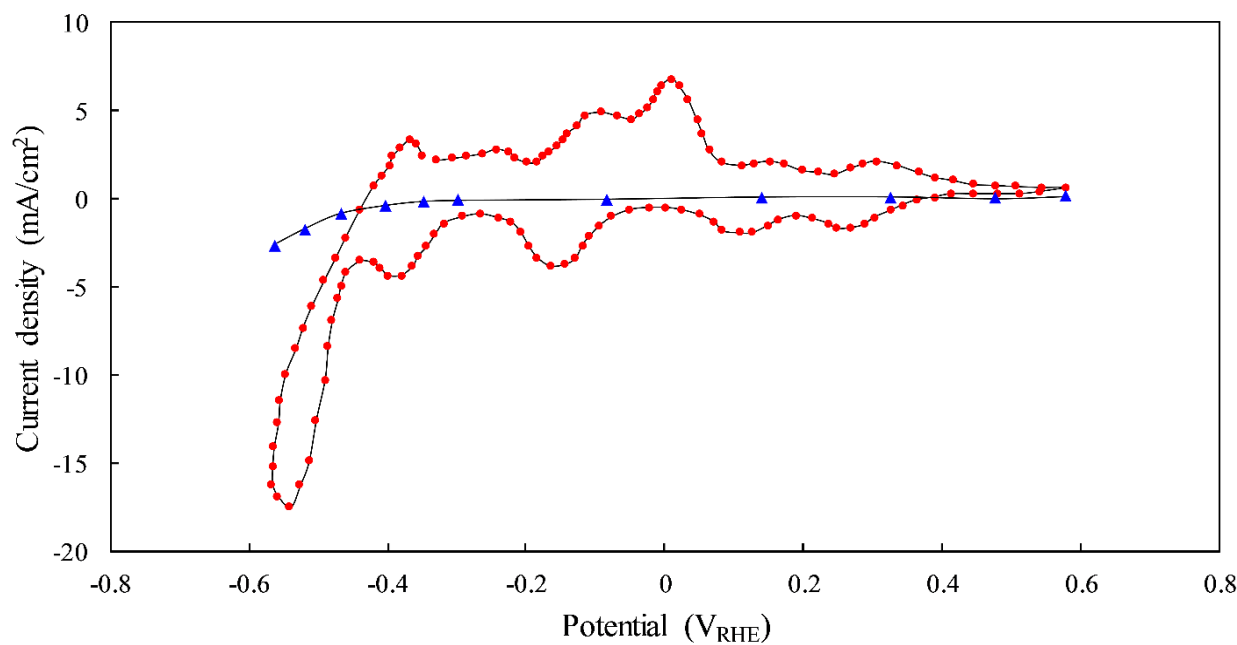

*Figure 13: Cyclic voltammogram (red circles) measured in an aqueous solution of 1M $H_2SO_4$ and 100 mM $Li_6[P_2W_{18}O_{62}]$, alongside a linear sweep voltammogram (blue triangles) measured in 1M $H_2SO_4$ aqueous solution without $Li_6[P_2W_{18}O_{62}]$. Both measurements were carried out using a glassy carbon working electrode. Plot reconstructed according to data from [19].*

Another interesting approach to electrochemical – chemical decoupled water splitting was presented by Amstutz *et al.,* employing a sophisticated modification of a redox-flow battery (RFB).[50] In their system, vanadium ($V^{+3}/V^{+2}$) and cerium ($Ce^{+4}/Ce^{+3}$) cations were used as soluble redox couples in the catholyte and anolyte of the RFB in its charging phase, respectively. Figure 14 presents a cyclic voltammogram measured in an aqueous solution of 1M $H_2SO_4$ and 50 mM vanadium(IV) (converted to vanadium(III)) sulfate using a graphite polymer rod as working electrode, displaying a redox wave centered around -0.3 $V_{RHE}$. Overlaid in Figure 14 is the corresponding LSV scan of the same graphite polymer rod in 1M $H_2SO_4$ without vanadium, showing that the HER current under these conditions is negligible up to -0.5 $V_{RHE}$. Thus, similarly to the STA redox couple, $V^{+3}$ can be reduced to $V^{+2}$ without concurrent hydrogen evolution at a graphite polymer rod electrode, yet it can potentially reduce water in the presence of a suitable catalyst.

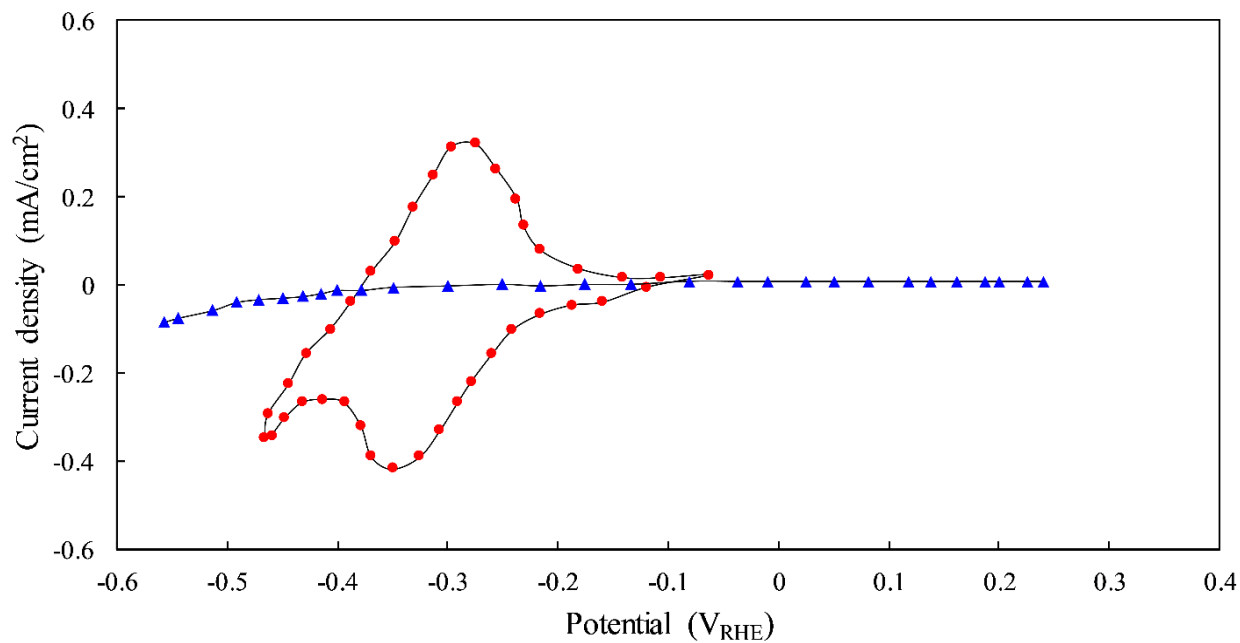

*Figure 14: Cyclic voltammograms measured in an aqueous solution of 1M $H_2SO_4$ with (red circles) and without (blue triangles) 50 mM vanadium(IV) sulfate (converted to vanadium(III) sulfate). Both measurements were carried out using a graphite polymer rod working electrode. Plots reconstructed according to data from* [50].

A complementary behavior is observed for the cerium ($Ce^{+4}/Ce^{+3}$) redox couple, as shown in Figure 15. Under the same conditions, the $Ce^{+4}/Ce^{+3}$ redox wave is centered around 1.4 $V_{RHE}$, whereas the OER current starts to rise above 1.5 $V_{RHE}$. Accordingly, $Ce^{+3}$ can be oxidized at a carbon electrode without oxygen evolution, but it could also potentially oxidize water in the presence of a suitable catalyst.

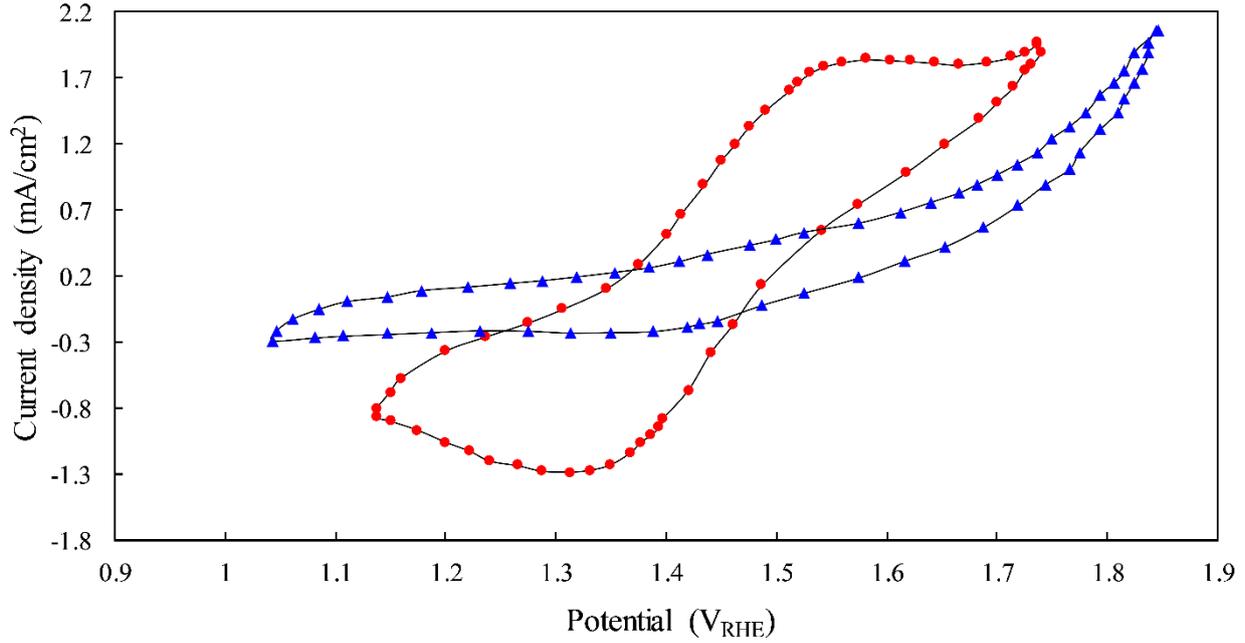

*Figure 15: Cyclic voltammograms measured in an aqueous solution of 1M $H_2SO_4$ with (red circles) and without (blue triangles) 50 mM cerium(III) sulfate. Both measurements were carried out using a graphite polymer rod working electrode. Plots reconstructed according to data from [50].*

Amstutz *et al.* demonstrated decoupled water splitting using their modified RFB operated in two steps, as illustrated in Figure 16. First, 50 ml aqueous solutions of 1M $H_2SO_4$ with 100 mM vanadium(III) sulfate (converted from vanadium(IV) sulfate) or cerium(III) sulfate were circulated through the RFB's catholyte and anolyte compartments, respectively. The battery was charged by reducing and oxidizing $V^{+3}$ and $Ce^{+3}$, respectively, at graphite felt electrodes. The reactions taking place during the charging step were:

$$\text{Anode: } Ce^{3+} \rightarrow Ce^{4+} + e^- \text{, } E^0 \cong 1.4 \text{ V}_{RHE} \quad \text{(Rxn 11)}$$

$$\text{Cathode: } V^{3+} + e^- \rightarrow V^{2+} \text{, } E^0 \cong -0.3 \text{ V}_{RHE} \quad \text{(Rxn 12)}.$$

This step was carried out by applying a constant current of 120 mA (current density of 60 mA $cm^{-2}$), with an average applied voltage of 2.5 V at a Faradaic efficiency of 94%. The charging step lasted 70 min, corresponding to a total charge of 504 C, or 5040 C $l^{-1}$.

In a completely electrolytic scheme, the battery was galvanically discharged, producing electricity, with a cell voltage of 0.7 V at a current of 120 mA (60 mAcm$^{-2}$):

$$\text{Cathode: } Ce^{4+} + e^- \rightarrow Ce^{3+}, \; E^0 \cong 1.4 \text{ V}_{RHE} \quad \text{(Rxn 11, reversed)}$$

$$\text{Anode: } V^{2+} \rightarrow V^{3+} + e^-, \; E^0 \cong -0.3 \text{ V}_{RHE} \quad \text{(Rxn 12, reversed)}.$$

In the alternative ECAC water splitting scheme, the charged electrolytes containing $Ce^{+4}$ and $V^{+2}$ were circulated through secondary cells, where they were brought into contact with catalytic particle beds of $RuO_2$ and $Mo_2C$, respectively, to accelerate their spontaneous chemical regeneration and gas evolution reactions:

$$4Ce^{4+} + 2H_2O \xrightarrow{RuO_2} 4Ce^{3+} + 4H^+ + O_2 \quad \text{(Rxn 13)}$$

$$2H^+ + 2V^{2+} \xrightarrow{Mo_2C} H_2 + 2V^{3+} \quad \text{(Rxn 14)}.$$

While almost all of the charge stored in the reduced $V^{+2}$ was restored following its chemical re-oxidation, with a reaction yield of (96 ± 4)%, the reaction yield for the $Ce^{+4}$ chemical reduction was only (78 ± 8)%. This could be attributed to a parasitic chemical reaction (such as $RuO_2$ corrosion by $Ce^{+4}$ cations),[51] or to an incomplete chemical reduction of the $Ce^{+4}$. The latter option seems to be highly likely, since the $Ce^{+4}$ reduction peak is situated at 1.3 V$_{RHE}$ (see Figure 15), only 70 mV above the OER standard potential. Considering the overpotential for the OER, even on an excellent OER catalysts such as $RuO_2$, the potential difference between the $Ce^{+4}$ reduction reaction and the OER, *i.e.*, the driving force for oxygen evolution, is quite small, and it continues to fall as more $Ce^{+4}$ is converted to $Ce^{+3}$. This deaccelerates the conversion rate until a complete halt in the spontaneous chemical reaction occurs once the potentials are equalized.

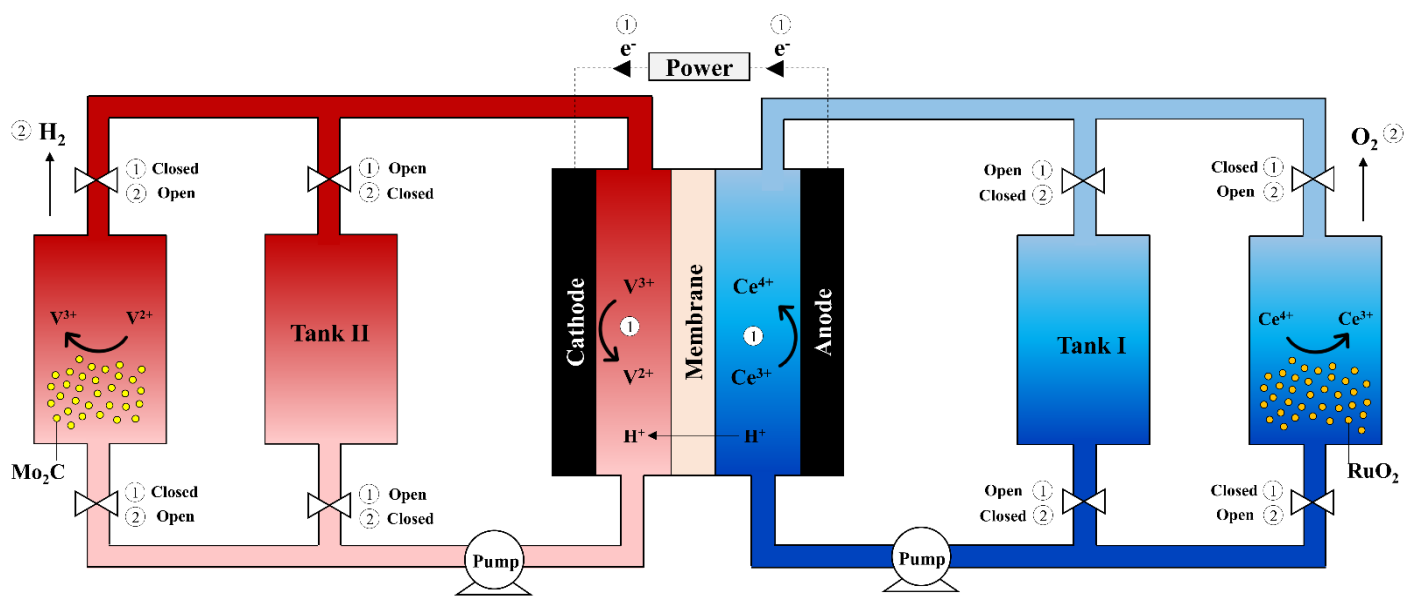

*Figure 16: Schematic illustration of an electrochemical – catalyst-activated chemical decoupled water splitting scheme with two soluble redox couples, cerium and vanadium, following the system proposed by Amstutz et al.[50] According to this scheme, water splitting is carried out in two steps. In the first step (step ①), cerium is oxidized ($Ce^{+3} \rightarrow Ce^{+4}$) at the anode of a modified redox flow battery (RFB), while vanadium is reduced ($V^{+3} \rightarrow V^{+2}$) at the cathode of the modified RFB. Then, in the second step (step ②), the solutions containing the oxidized cerium ($Ce^{+4}$) and the reduced vanadium ($V^{+2}$) are circulated to two other cells containing $RuO_2$ and $Mo_2C$ catalysts, respectively. Once in contact with the catalysts, the $Ce^{+4}$ and $V^{+2}$ solutions undergo spontaneous chemical reduction and oxidation to their initial states ($Ce^{+3}$ and $V^{+3}$) while releasing oxygen and hydrogen, respectively.*

**Electrochemical – chemical water splitting with solid redox electrodes**

Nickel (oxy)hydroxide electrodes were already discussed previously as auxiliary redox electrodes for decoupled electrolytic water splitting in alkaline solutions. As mentioned, the standard potential of the $Ni(OH)_2$/NiOOH redox reaction is higher than that of the OER. Therefore, its oxidized form, NiOOH, can oxidize water and produce oxygen, in a chemical reaction that reduces the NiOOH to $Ni(OH)_2$, in an analogous manner to the spontaneous chemical re-oxidation of the STA redox couple discussed previously. Indeed, chemical reduction of NiOOH with concurrent oxygen evolution is an unintentional (and undesired) form of self-discharge in rechargeable (secondary) alkaline batteries such as Ni-$H_2$ and Ni-MH batteries.[52,53] This spontaneous reaction is driven by the difference between the electrochemical potentials (*i.e.*, free energies) of the oxidized NiOOH electrode and the OER. At ambient temperatures this reaction is hindered kinetically, enabling alkaline batteries to operate thousands of charge-discharge cycles without parasitic oxygen evolution that causes swelling and eventually battery failure.[39,54] However, at elevated temperatures the electrode's self-discharge reaction is accelerated, following the Arrhenius equation.[30,55] The self-discharge reaction can be written as:

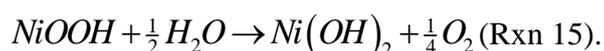
$$NiOOH + \tfrac{1}{2} H_2O \rightarrow Ni(OH)_2 + \tfrac{1}{4} O_2 \text{ (Rxn 15)}.$$

Dotan *et al.* were the first to report that the self-discharge reaction of NiOOH to Ni(OH)$_2$ (Rxn 15) can be employed in an electrochemical – thermally-activated chemical (ETAC) decoupled water splitting cycle, using a Ni(OH)$_2$ electrode as the primary anode and a HER cathode in a cell with alkaline aqueous electrolyte.[30] In this scheme, the Ni(OH)$_2$ anode acts as an electron-coupled hydroxide buffer (ECHB) that undergoes electrochemical oxidation to NiOOH during the hydrogen generation step, followed by spontaneous chemical reduction of the anode back to Ni(OH)$_2$ while producing oxygen according to Rxn 15. Figure 17 shows a cyclic voltammogram measured at ambient temperature (~25°C) in 5M KOH aqueous electrolyte with a cobalt-doped Ni(OH)$_2$ working electrode, displaying one redox wave centered around 1.4 V$_{RHE}$. Overlaid in this graph are the steady-state current – potential plots for the same electrode, in the same electrolyte, measured at ambient temperature (▲) and at 95°C (●). At ambient temperature, the steady-state OER current density is less than 1 mA cm$^{-2}$ near the electrode's redox potential (~1.4 V$_{RHE}$). Thus, although the chemical reduction of the oxidized electrode (Rxn 15) is a spontaneous reaction, it occurs at a very slow rate at ambient temperature. However, the reaction rate can be significantly enhanced by raising the temperature, as can be seen in the respective curves (●) in Figure 17.

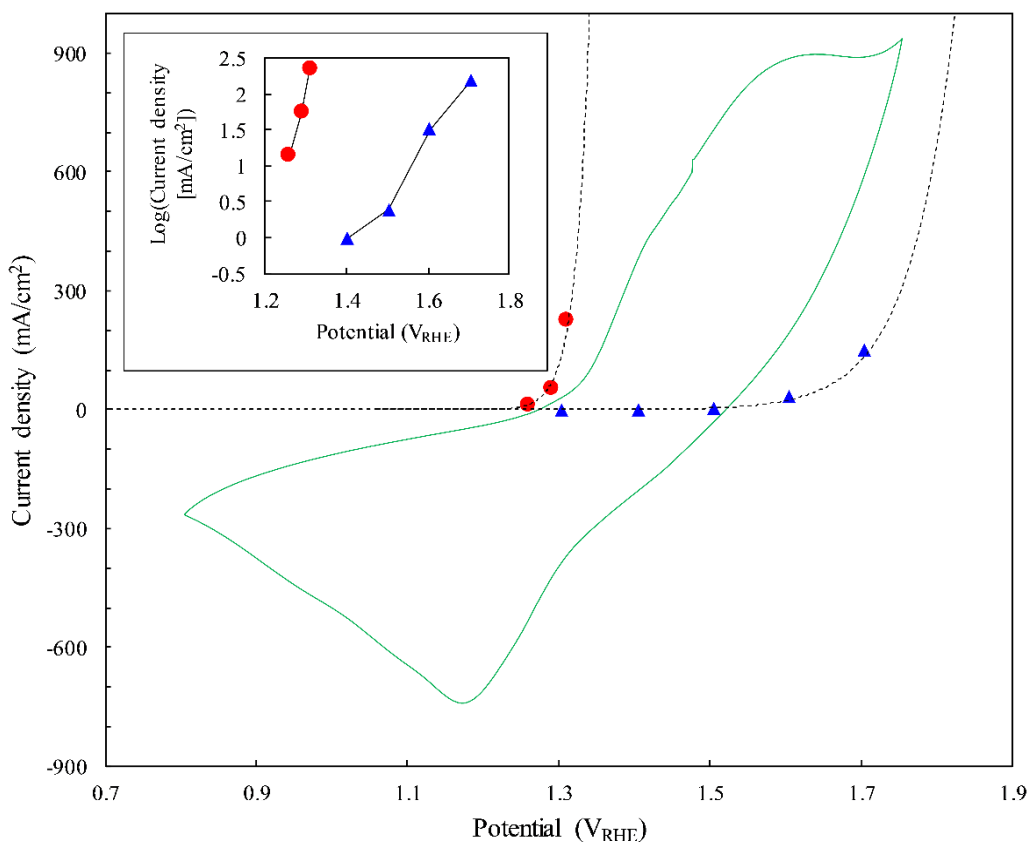

*Figure 17: Cyclic voltammogram measured at ambient temperature (~25°C) in 5M KOH aqueous electrolyte with a cobalt-doped Ni(OH)₂ working electrode (green line), alongside steady-state current – potential plots for the same electrode in the same electrolyte measured at ambient temperature (blue triangles) and at 95°C (red circles). The dashed lines are extrapolated exponential curves. The inset shows the Tafel plots for the steady state polarization at ambient temperature and 95°C (same color coding). All plots were reconstructed according to data from [30].*

In the ETAC decoupled water splitting cycle reported by Dotan *et al.*,[30] a cobalt-doped Ni(OH)₂ electrode was first connected as the anode of a two-electrode electrochemical cell with a platinized nickel-coated stainless steel cathode and 5M KOH aqueous electrolyte. In the first step, the cell was operated at ambient temperature (~25°C) at a constant current density of 50 mA cm⁻² for 100 s. The cathodic and anodic reactions during this step were:

Cathode: $H_2O + e^- \rightarrow \tfrac{1}{2} H_2 + OH^-$ , $E^0 = 0$ $V_{RHE}$ (Rxn 6)

Anode: $Ni_{0.9}Co_{0.1}(OH)_2 + OH^- \rightarrow Ni_{0.9}Co_{0.1}OOH + H_2O + e^-$ , $E^0 \cong 1.35$ $V_{RHE}$ (Rxn 5, modified).

The reversible potential of the cobalt-doped Ni(OH)$_2$ electrode was ~1.35 V$_{RHE}$, slightly lower than the standard potential of the Ni(OH)$_2$/NiOOH redox couple (~1.42 V$_{RHE}$)[45] due to the addition of cobalt, which shifts the redox potential cathodically.[37,56]

After charging the cobalt-doped Ni(OH)$_2$ anode to a threshold charge of 5 C cm$^{-2}$, the power was switched off (open circuit). Then, in the second step of the ETAC cycle the charged cobalt-doped NiOOH anode was exposed to a hot (95°C) 5M KOH aqueous electrolyte for 100 s, whereupon the anode was spontaneously reduced back to its original state (Ni(OH)$_2$) while releasing oxygen according to the following chemical reaction:

$$Ni_{0.9}Co_{0.1}OOH + \tfrac{1}{2}H_2O \rightarrow Ni_{0.9}Co_{0.1}(OH)_2 + \tfrac{1}{4}O_2 \text{ (Rxn 15, modified)}.$$

At the end of the second step, the anode was regenerated back to its initial state. The overall reaction, that is the sum of Rxns 5 (modified), 6 and 15 (modified), is the water splitting reaction: H$_2$O → H$_2$ + ½O$_2$ (Rxn 4). The basic steps in the ETAC water splitting cycle are illustrated in Figure 18A, and Figure 18B illustrates a batch-mode operation scheme in a cell with stationary electrodes and alternating circulation of cold and hot electrolytes through the cell.

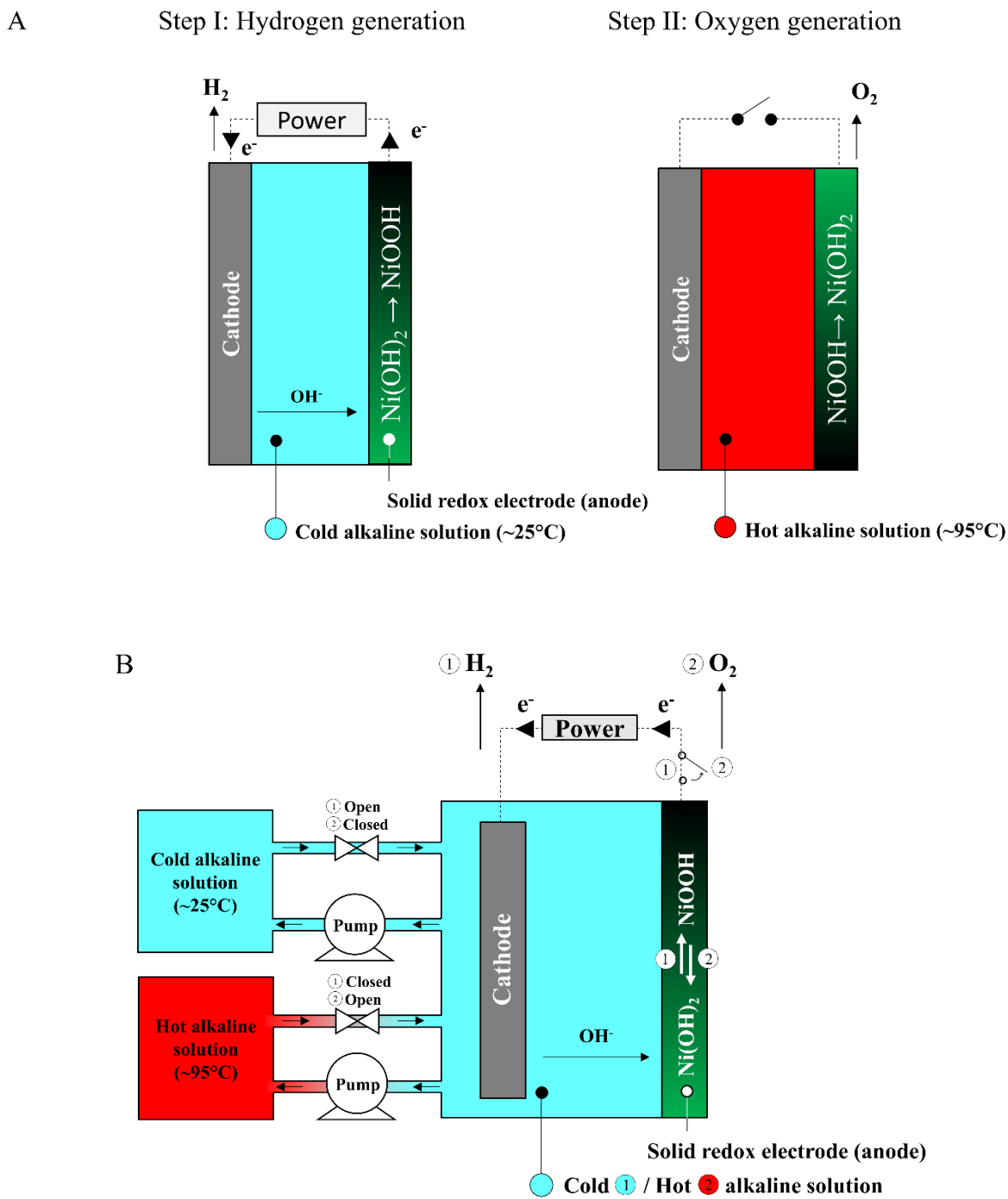

*Figure 18: Schematic illustration of the ETAC decoupled water splitting cycle proposed by Dotan et al.[30] The electrochemical hydrogen generation step (Step I, left), is carried out in a cold (25°C) alkaline solution. During this step, the $Ni(OH)_2$ anode is oxidized while hydrogen is evolved at the cathode. Then, the chemical oxygen generation step (Step II, right), is carried out in a hot (95°C) alkaline solution at open circuit. During this step, the oxidized NiOOH anode is spontaneously reduced back to its initial state ($Ni(OH)_2$), in a chemical reaction with water (Rxn 15) that evolves oxygen. (A) Illustration of the basic steps. (B) Illustration of a batch-mode operation scheme in a cell with stationary electrodes and alternating circulation of the cold and hot electrolytes. In the first step ①, a cold (25°C) electrolyte (light grey) is circulated though the cell and the anode and cathode are connected to a power source. Then, in the second step ②, the power is switched off and a hot (95°C) electrolyte (dark grey) is circulated through the cell.*

In the proof-of-concept experiment reported by Dotan *et al.*,[30] ten stable ETAC water splitting cycles were demonstrated by alternately placing the cobalt-doped nickel (oxy)hydroxide anode in cold and hot alkaline solutions. This was done manually for demonstration purposes, but it was noted that a real ETAC water splitting system would have cells with stationary electrodes, and it would be operated in a swing mode by alternately circulating the cold and hot solutions through the cells, as illustrated in Figure 18B. Each electrochemical (hydrogen generation) step was carried out at ambient temperature in 5M KOH aqueous electrolyte at a current density of 50 mA cm$^{-2}$ for 100 s, and each chemical (oxygen generation) step was carried out in a hot (95°C) 5M KOH aqueous electrolyte for 100 s. In each hydrogen generation step, the cell voltage increased progressively, as shown in the inset in Figure 19. The increase in the cell voltage during each step followed the increase in the anode's potential as it was charged during the step. Overall, the anode's potential varied between 1.37 and 1.45 V$_{RHE}$, with an average potential of 1.42 V$_{RHE}$. The overall cell voltage in the hydrogen generation steps varied between 1.44 and 1.56 V, with an average voltage of 1.5 V, corresponding to an average voltage efficiency of $1.48\text{V}/1.5\text{V} = 98.7\%_{HHV}$. In another experiment, 100 cycles were also demonstrated, displaying similar behavior with no signs of performance degradation.[30]

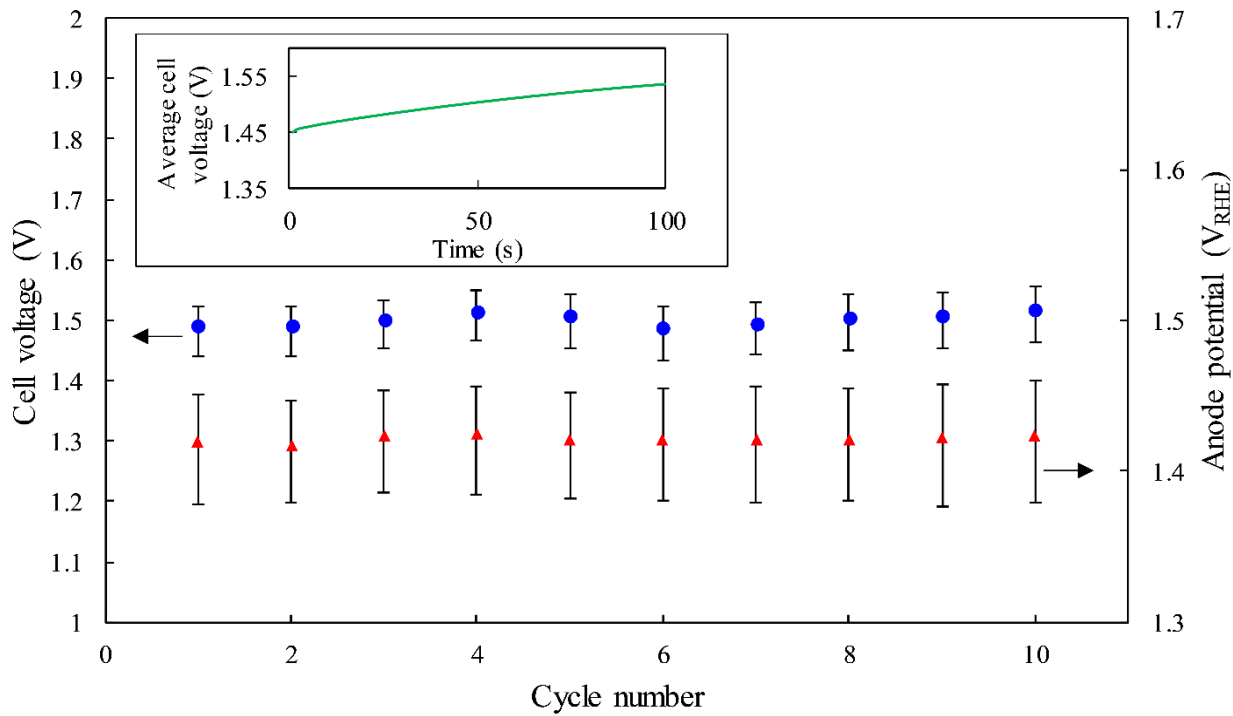

*Figure 19: The average cell voltage (blue circles) and range (bars), and the corresponding anode potential (red triangles) and range (bars), in ten ETAC water splitting cycles carried out with a cobalt-doped nickel (oxy)hydroxide redox electrode. The inset presents the voltage evolution throughout a cycle, averaged over all ten cycles. All plots were reconstructed according to data from [30].*

Since hydrogen is the desired product, the duration of other processes besides the hydrogen generation step (*e.g.*, the oxygen generation step) should be minimized in order to maximize the hydrogen production throughput. In the proof-of-concept experiment reported by Dotan *et al.*,[30] the oxygen generation and hydrogen generation steps had an equal duration of 100 s. While the hydrogen is generated electrochemically and therefore its production rate is directly controlled by the applied current, the oxygen is generated in a spontaneous chemical reaction and therefore its production rate cannot be controlled directly. Nevertheless, the anode regeneration rate can be manipulated by temperature to hinder oxygen generation in the first step (hydrogen generation) that occurs at near ambient temperature, and to accelerate oxygen generation in the second step that occurs at high temperature (~95°C), as discussed above. Thus, temperature is an important control parameter in the ETAC water splitting process, enabling decelerating or accelerating oxygen generation in the first and second steps, respectively, in a similar manner to temperature swing absorption (TSA) processes.

Although kinetically inhibited, oxygen may still evolve parasitically at the charging Ni(OH)$_2$ anode even at ambient temperature, especially when it is overcharged.[57] Figure 20 shows a typical galvanostatic charging (oxidation) curve for a commercial nickel hydroxide anode (at ambient temperature).[25] After a certain SOC (in this case ~50%), the anode's potential increases sharply from 1.4 to 1.5 $V_{RHE}$. Around this point, the charging potential plateau shifts towards the OER plateau, and oxygen starts to evolve at the overcharging anode (see inset in Figure 20). To avoid co-generation of hydrogen and oxygen, which might be detrimental since there is no separator between the cathode and anode (see Figure 18), the first step (hydrogen generation) must be stopped before the anode is overcharged. This can be achieved by setting a potential limit to the charging anode, a SOC limit, a voltage limit (between the anode and cathode), or by monitoring the concentration of dissolved oxygen during the charging (first) step.

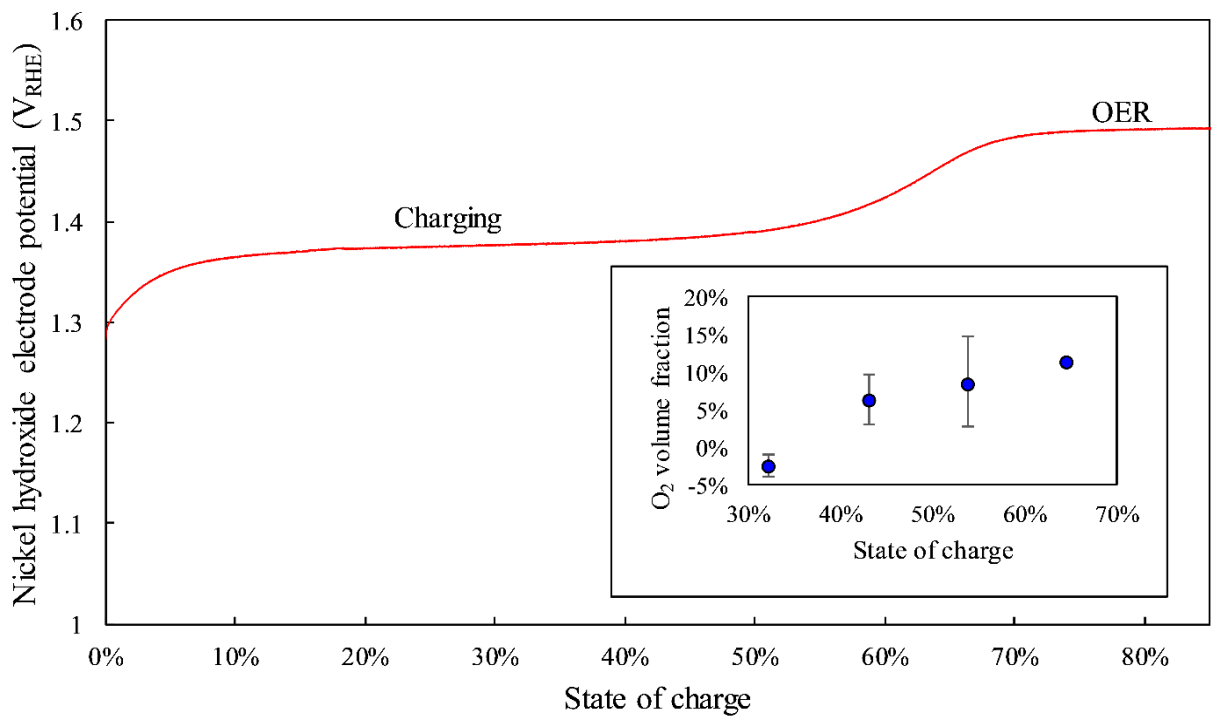

*Figure 20: A galvanostatic charging curve of a commercial nickel hydroxide anode taken in 1M NaOH aqueous electrolyte at ambient temperature. The inset shows the corresponding oxygen volume fraction (measured by gas chromatography) of the total effluent gas volume as a function of the SOC. All plots were reconstructed according to data from [25].*

The ETAC process presents several key advantages as a prospective new technology for decoupled water splitting. First, decoupled membraneless operation could pave the way to high-pressure hydrogen production, avoiding $H_2$ / $O_2$ leakage through the separator at high pressures that limits the operation pressure of conventional alkaline electrolyzers.[58] Second, the process could principally be operated without precious metal catalysts, since it is carried out in an alkaline solution where stable nickel-based catalysts are available.[59] Third, the anode charging and spontaneous chemical regeneration reactions (Rxns 5 and 15) replace the electrochemical oxygen evolution reaction (Rxn 7), thereby reducing the overpotential loss for water oxidation (oxygen generation) which typically contributes the largest energy loss in alkaline and PEM electrolyzers.[5] To demonstrate this, Figure 21 compares CV and steady-state Tafel plots (inset) of the cobalt-doped nickel (oxy)hydroxide anode from the ETAC water splitting proof-of-concept experiment presented in Figure 19 and a similar anode functionalized with NiFe layered-double hydroxide (LDH) catalyst, a state-of-the-art OER catalyst in alkaline solutions.[60] All the measurements were carried out in 5M KOH aqueous electrolyte at ambient temperature. Although the NiFe LDH functionalized anode outperforms the cobalt-doped nickel (oxy)hydroxide anode in the steady-state OER current, as can be seen by in the respective Tafel plots in the inset of Figure 21, the oxidation (charging) reaction of the cobalt-doped nickel (oxy)hydroxide anode precedes the OER at the NiFe LDH functionalized anode. Dotan *et al.* demonstrated stable ETAC water splitting cycles at a current density of 50 mA cm$^{-2}$ with an average anode potential of 1.42 V$_{RHE}$, whereas the NiFe LDH functionalized anode required a potential of 1.53 V$_{RHE}$ to achieve the same current density.

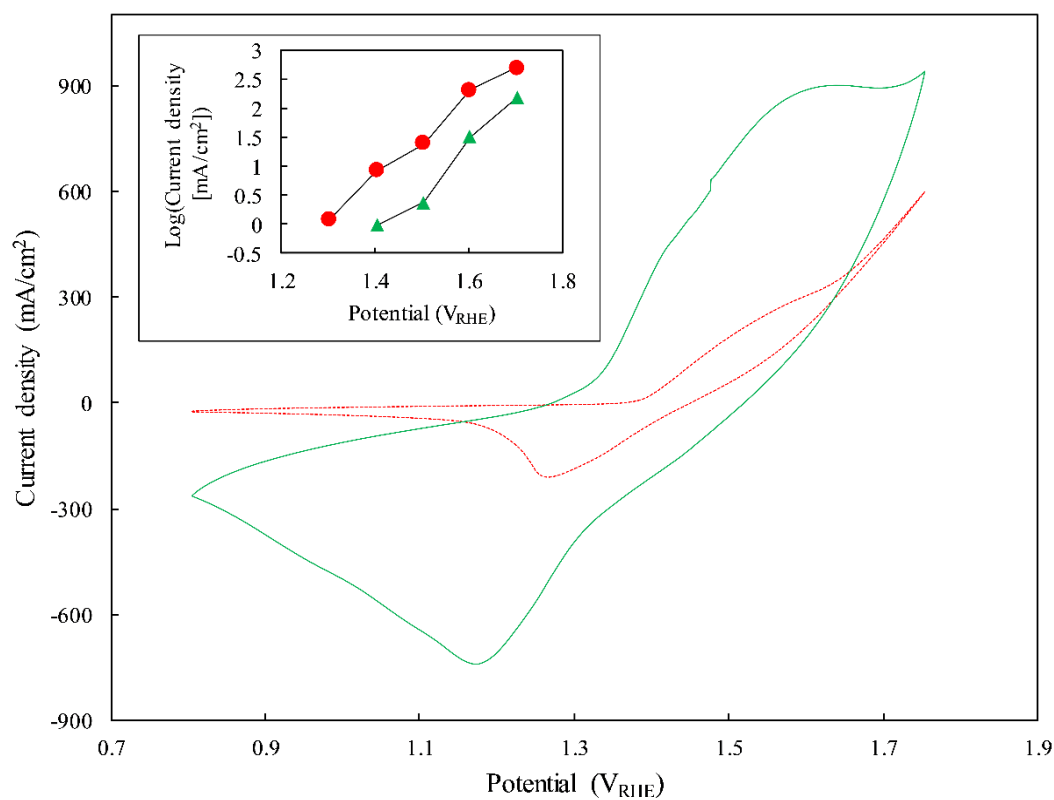

*Figure 21: Cyclic voltammograms measured at ambient temperature in 5M KOH aqueous electrolyte with a cobalt-doped nickel (oxy)hydroxide anode (solid green line) and a NiFe layered double hydroxide (LDH) anode (dashed red line). The inset shows the corresponding Tafel plots for steady-state polarization of the cobalt-doped nickel (oxy)hydroxide anode (green triangles) and the NiFe LDH anode (red circles). All plots were constructed according to data from [30].*

The ETAC water splitting process is patented,[23] and commercial realization efforts are underway by the Israeli startup company H$_2$Pro.[61] Figure 22A shows a schematic illustration of the operation mode of a continuous hydrogen generator using multiple cells operated in a cyclic configuration with electrolyte circulation. In this swing mode operation, two cells are operated concurrently, alternating between the hydrogen and oxygen generation steps, such that Cell A generates hydrogen at near ambient temperature while Cell B generates oxygen at high temperature (~95°C), and vice versa. The electrodes in both cells are stationary, and the cold and hot alkaline solutions are circulated alternately through the cells to facilitate hydrogen and oxygen evolution, respectively. The cycle that each cell undergoes is illustrated in Figure 22B.

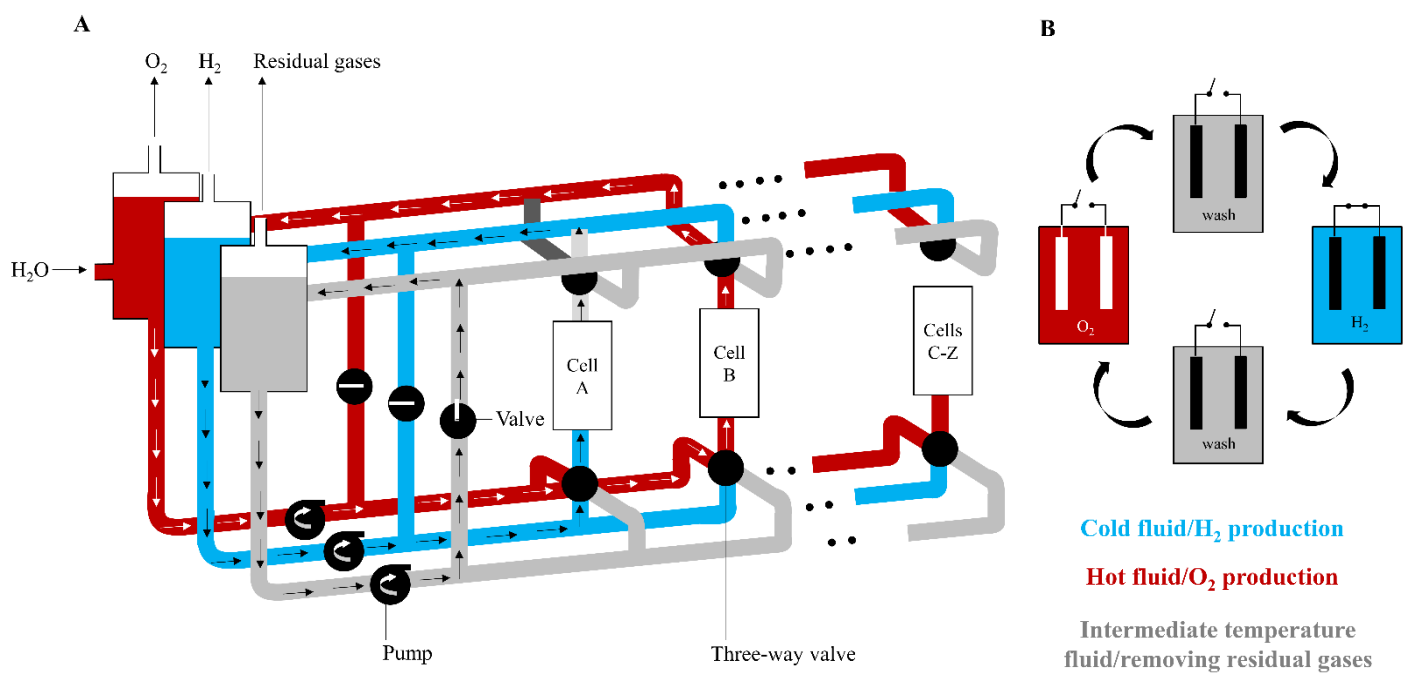

*Figure 22: Schematic illustration of the operation mode of a multi-cell system for continuous hydrogen generation using the ETAC water splitting process proposed by Dotan et al.[30] (A) Scheme of the system and the electrolyte flow during operation. In this snapshot,* Cell A *is the hydrogen generation cell and* Cell B *is the oxygen generation cell, and the two cells operate concurrently while a cold electrolyte (blue) is circulated through* Cell A *and a hot electrolyte (red) is circulated through* Cell B. *Between cycles, an intermediate temperature electrolyte (grey) can be circulated through the cell to wash any residual gases.(B) Scheme of a single-cell cycle, showing the different stages according to the different fluid in the cell.*

During each step in the cycle, the electrolyte that is pumped from the reservoir to the cell moves the gas bubbles back into the reservoir, which serves as a phase separator. Each separator releases the gasses from the top as shown in Figure 22A, and the gas-free electrolyte is re-circulated. In between the hydrogen and oxygen generation steps, there is an intermediate washing step, which moves the residual electrolyte in the cells to the respective reservoir. Minimizing heat losses during this cycle is crucial; In this task the reservoir with the washing electrolyte acts as heat ballast, providing initial heat to the cells, following the hydrogen generation step, and providing an initial quench following the oxygen generation step. Besides this measure, the hot reservoir and wash reservoir are well insulated to minimize heat losses to the environment. In addition, the production cells have internal insulation to minimize heat losses to their external shells. Finally, the oxygen generation reaction (Rxn 15) is spontaneous and exothermic, therefore with a proper design of the electrodes and cells, the heat supplied by this step suffices to compensate for the heating of the electrodes and inner cell walls during the cycle. A prototype ETAC water splitting demonstration system is presented in Figure 23.

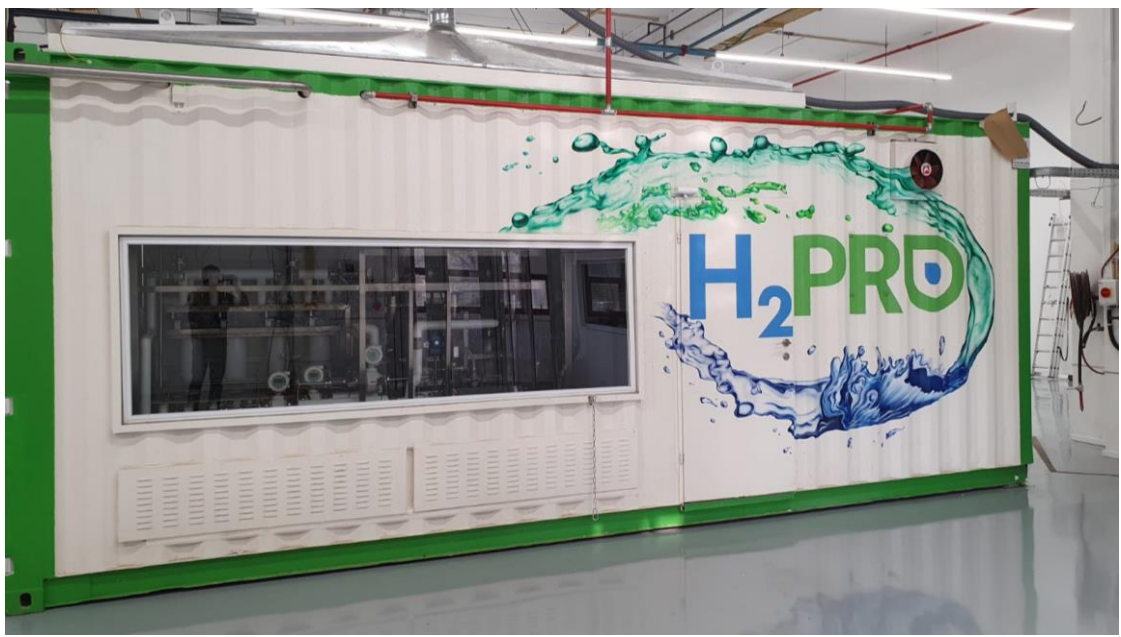

*Figure 23: Prototype ETAC water splitting demonstration system. Courtesy of H$_2$Pro.*[61]

A different approach for electrochemical - chemical decoupled water splitting is pursued by the French startup company Ergosup,[32] offering another path to high-pressure hydrogen production. This patented approach[31] is based on zinc (Zn) cathodic deposition from zinc sulfate (ZnSO4) solution with concomitant oxygen evolution in the first step (an electrochemical step), followed by spontaneous dissolution of the zinc coating in sulfuric acid (H2SO4) solution with concomitant hydrogen evolution in the second step (a chemical step that proceeds without applied bias). Figure 24 illustrates the Ergosup water splitting cycle. The first step is carried out electrolytically, generating oxygen at the anode and zinc coating at the cathode:

$$\text{Anode: } H_2O_{(aq)} \rightarrow \tfrac{1}{2}O_{2(g)} + 2H^+_{(aq)} + 2e^- \quad E^0 = 1.23 V_{RHE} \quad \text{(Rxn 1)}$$

$$ZnSO_{4(aq)} + 2H^+_{(aq)} \rightarrow H_2SO_{4(aq)} + Zn^{2+}_{(aq)} \quad \text{(Rxn 16)}$$

$$\text{Cathode: } Zn^{2+}_{(aq)} + 2e^- \rightarrow Zn^0_{(s)} \quad E^0 = -0.76 V_{RHE} \quad \text{(Rxn 17)}.$$

The sum of Rxns 1, 16 and 17 yields the following reaction:

$$ZnSO_{4(aq)} + H_2O_{(aq)} \rightarrow Zn^0_{(s)} + H_2SO_{4(aq)} + \tfrac{1}{2}O_{2(g)} \quad \text{(Rxn 18)},$$

which requires applying a voltage greater than 1.99 V (the reversible cell voltage) to proceed. This reaction is widely used for the production of zinc from zinc sulfate solutions in a process called zinc electrowinning.[62] The second step in Ergosup's decoupled water splitting process is a spontaneous chemical step wherein the zinc coating ($Zn^0_{(s)}$) from the first step is attacked by the acid, re-dissolving into the electrolyte while evolving hydrogen:

$$Zn^0_{(s)} + H_2SO_{4(aq)} \rightarrow ZnSO_{4(aq)} + H_{2(g)} \quad \text{(Rxn 19)}.$$

Here, the chemical step takes place immediately as soon as the power is switched off, and it does not require a catalyst or temperature swing as in the other electrochemical – chemical cycles described previously.

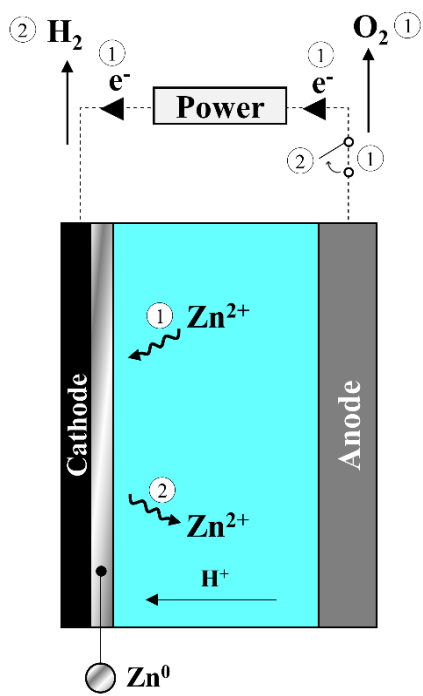

*Figure 24: Schematic illustration of the electrochemical – chemical decoupled water splitting process proposed by Ergosup.[31] According to this scheme water splitting is carried out in two steps. In the first step ①, water is oxidized at the anode while dissolved zinc ions ($Zn^{+2}_{(aq)}$) are reduced electrochemically and deposited at the cathode . Then, in the second step ②, the power is switched off (open circuit) and the deposited zinc layer ($Zn^0_{(s)}$) at the cathode undergoes spontaneous chemical dissolution while releasing hydrogen.*

The main advantage of the Ergosup process is that it enables high-pressure hydrogen production in a compact hydrogen generator. In 2019, Ergosup reported the first installation of their *HiRiS* laboratory hydrogen generator, featured in Figure 25.[63] The system is capable of producing 0.7 to 4 kg of hydrogen per week at a pressure of 100 bar. However, the Ergosup process has a high energy consumption, with a reversible cell voltage for the zinc electrowinning process (Rxns. 1, 16 and 17) of 1.23 $V_{RHE}$ – (–0.76 $V_{RHE}$) = 1.99 V. Under polarization, a cell voltage of 2.57 V was applied at a current density of 40 mA cm$^{-2}$,[31] corresponding to a voltage efficiency of 57.6%$_{HHV}$, considerably lower than state-of-the-art alkaline electrolyzers,[64] and much lower than the ETAC water splitting process where a voltage efficiency of 98.7%$_{HHV}$ was reported at a similar current density (50 mA cm$^{-2}$).[30]

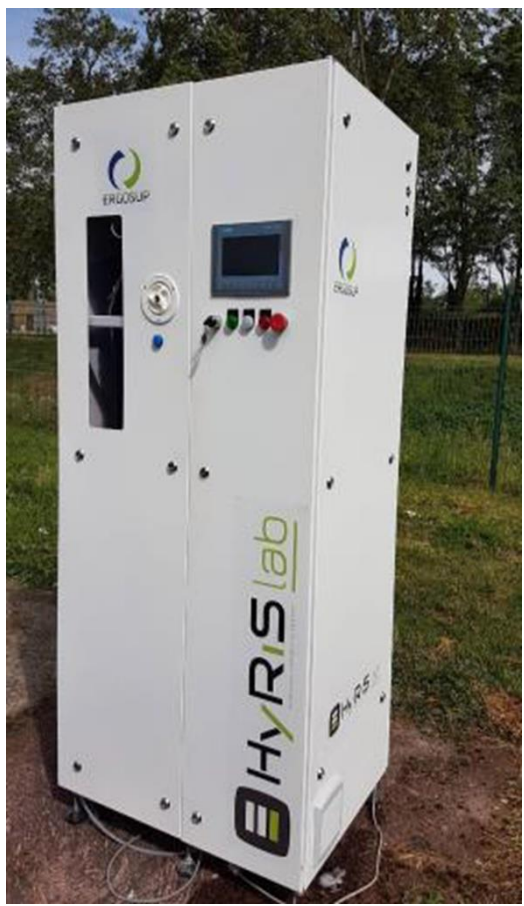

*Figure 25: Photograph of Ergosup's HiRiS laboratory hydrogen generator. Courtesy of Ergosup.*

**Decoupled photoelectrochemical and photocatalytic water splitting**

Photoelectrochemical (PEC) water splitting has been extensively studied over the past few decades, as an elegant route for direct conversion of solar energy to green hydrogen.[65] Theoretically, photovoltaic-photoelectrochemical (PV-PEC) tandem cells with ideal photoelectrodes could surpass the efficiency of PV-electrolysis systems that use the same PV cells,[66] because of the enhanced solar energy conversion efficiency of multi-junction tandem cells with respect to single-junction cells.[67] So far, however, the performance of photoelectrodes for PEC solar water splitting falls short of the minimum criteria necessary to compete with PV-electrolysis for efficiency and stability.[68] Even if the performance of new photoelectrodes was to be dramatically improved by R&D breakthroughs, the practical realization of large-scale PEC hydrogen production plants remains a challenge.

Due to the diffuse nature of natural sunlight, an immense area must be covered by PEC cells to produce hydrogen at a commercial scale. For instance, PEC cells operating 6 h a day at a photocurrent density of 10 mA cm$^{-2}$, a challenging benchmark level that so far has been demonstrated only with multi-junction tandem cells comprising of intrinsically unstable photoelectrode materials such as GaInP/GaAs and AlGaAs/GaInAs,[69,70] the net area of PEC cells required in order to produce 1000 m$^3$ (81.3 kg) H$_2$ day$^{-1}$ would be about 4,000 m$^2$. Assuming 10×10 cm$^2$ PEC cells, similar to the typical size of commercial PV cells, this means 400,000 PEC cells that produce hydrogen and oxygen concurrently in the same cell, all of which would have to be hermetically sealed and connected to a gas piping manifold to collect and transport the hydrogen gas to a central storage and distribution facility. Additionally, as any other hydrogen production plant, this array of PEC cells would have to be monitored in accordance with hydrogen safety guidelines,[71] keeping in mind that the safety issues related to partial-load operation will likely be exacerbated at the low-current densities characteristic of PEC cells. These technical issues present formidable challenges towards the realization of large-scale PEC water splitting for solar hydrogen production.

A disruptive approach to address these challenges is to decouple the hydrogen and oxygen evolution reactions such that one reaction occurs in one place and the other in another place. Specifically, if the oxygen evolution occurs in the PEC cell and the hydrogen evolution occurs elsewhere in an electrolytic cell that does not require exposure to sunlight, this may pave the way to PEC water splitting plants that produce hydrogen in a central location (say, at the edge of the solar field) whereas oxygen is produced in the PEC cells within the solar field, as shown in the conceptual illustration presented in Figure 26.

Namely, this can enable centralized hydrogen production in a separate compact vessel while the oxygen is produced in many PEC cells distributed in the solar field.

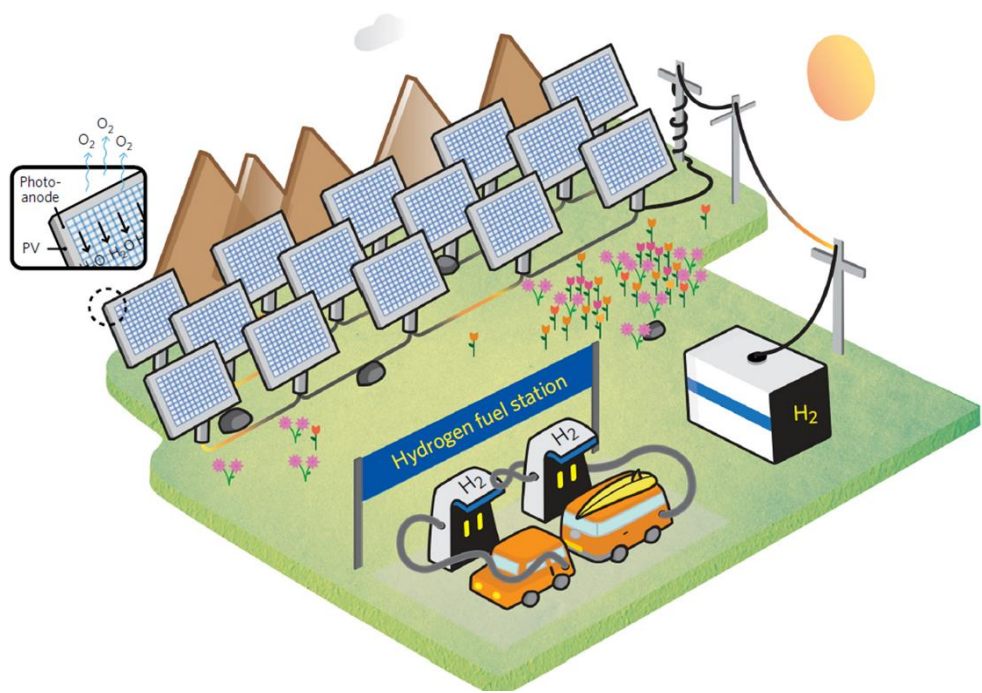

*Figure 26: Conceptual illustration of a decoupled PEC water splitting plant, with distributed PEC cells producing oxygen at the solar field and a central hydrogen generator at the edge of the field, nearby a hydrogen refueling station. Reproduced from Landman et al. [25] with permission (Copyright © 2017, Springer Nature).*

Following this logic, a prototype device for decoupled PEC water splitting in separate hydrogen and oxygen cells was demonstrated by Landman *et al.*,[34] using the $Ni(OH)_2$/NiOOH hydroxide-mediating auxiliary redox electrodes discussed previously. The system comprised of two cells connected in series, as shown in Figure 27.

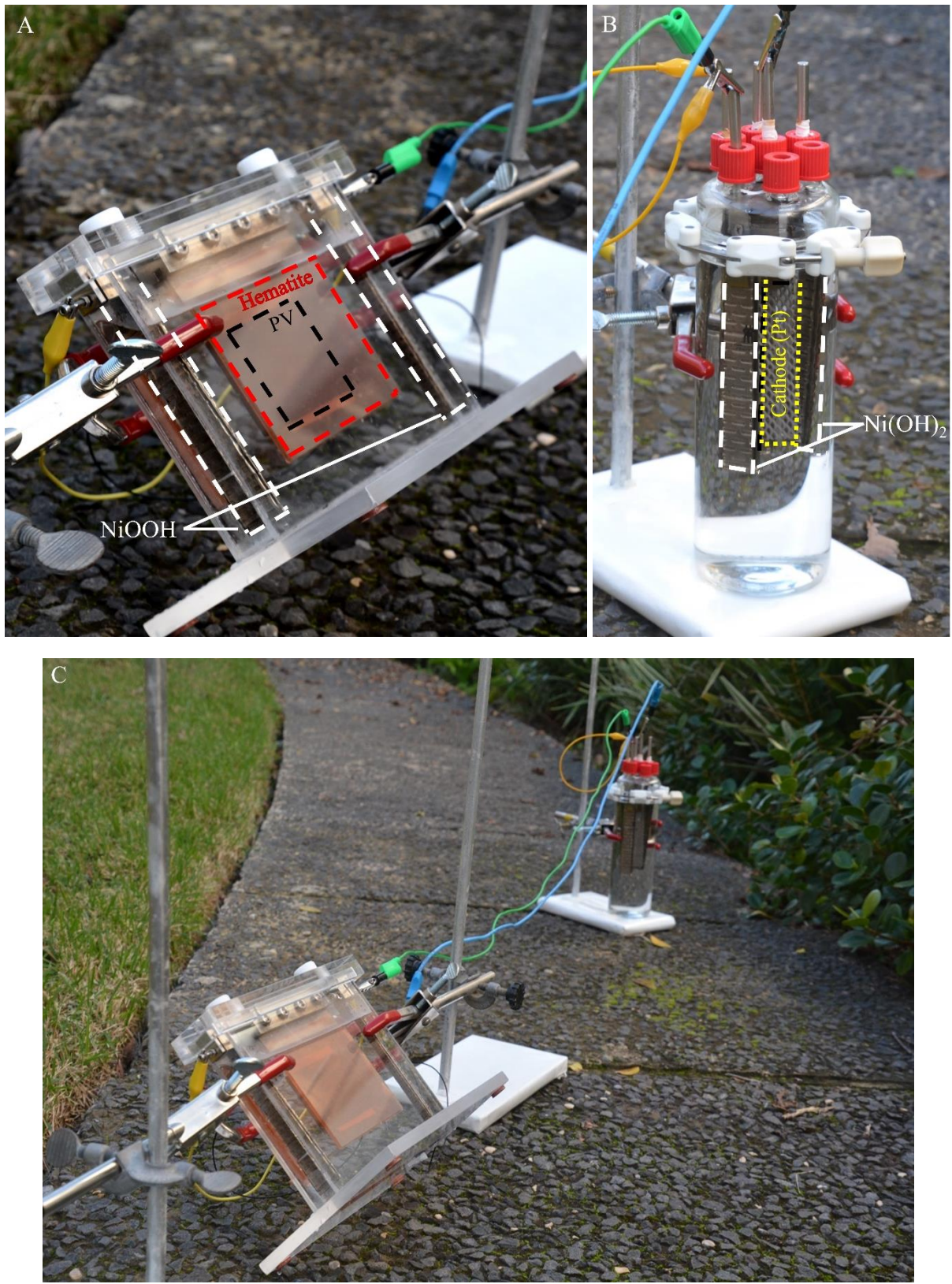

*Figure 27: Photographs of the decoupled PEC water splitting device presented by Landman et al.[34] (A) The oxygen generation cell; (B) The hydrogen generation cell; (C) The whole device with the two cells connected electrically by conducting wires. The oxygen cell contains a 100 cm$^2$ hematite photoanode on top of a 50 cm$^2$ Si PV module with two NiOOH auxiliary redox electrodes. The hydrogen cell contains a platinized Ti-mesh cathode and two Ni(OH)$_2$ auxiliary redox electrodes.*

The oxygen cell in the system presented by Landman et al.[34] was a PEC-PV tandem cell, wherein a 100 cm$^2$ hematite (α-Fe$_2$O$_3$) photoanode was dipped in 1M NaOH aqueous solution. To supply the additional bias required for water splitting, a Si PV module was connected in tandem to and behind the PEC cell. Instead of a hydrogen evolving cathode, the cell was equipped with two battery-grade NiOOH auxiliary redox electrodes. By contrast, the hydrogen cell was an electrochemical cell that contained a platinized Ti mesh cathode and two Ni(OH)$_2$ auxiliary redox electrodes in 1M NaOH aqueous solution. The two cells were connected in series, as illustrated in Figure 8C, wherein the PV module was electrically connected to the photoanode in the oxygen cell and to the cathode in the hydrogen cell, and the auxiliary redox electrodes within the two cells were connected in short circuit to each other. Under illumination, oxygen evolved at the hematite photoanode while the NiOOH auxiliary redox electrodes were reduced to Ni(OH)$_2$ in the oxygen cell. At the same time, hydrogen evolved at the cathode in the (separate) hydrogen cell, while the Ni(OH)$_2$ auxiliary redox electrodes in that cell were oxidized to NiOOH.

No hydrogen was evolved in the illuminated PEC oxygen cell and no oxygen was evolved in the electrolytic hydrogen cell, as discussed previously, eliminating the need for membranes in both cells. Moreover, the oxygen that was produced in the illuminated PEC cell could simply be released to the atmosphere, eliminating the need for hermetic sealing as well as the complex gas piping manifold for the collection of hydrogen from the PEC cells in the solar field. Finally, the electrical connection between the two cells enables their separation over quite large distances, such that a central compact hydrogen production unit could, in principle, be placed at the edge of the solar field, as illustrated in Figure 26.

Decoupling of the photoelectrochemical OER and the electrochemical HER therefore offers an elegant solution to the technical challenges discussed above. Nevertheless, the integration of the Ni(OH)$_2$/NiOOH auxiliary redox electrodes introduces some new challenges. The main challenge is swapping these electrodes when they reach their full operational capacity, as illustrated in Figure 8C. The spent NiOOH electrodes must be periodically collected from all the PEC cells in the solar field and be swapped with the spent Ni(OH)$_2$ electrodes from the hydrogen cell, a task that adds complexity and will undoubtedly increase operational costs. This challenge can be addressed by placing a pack of extra auxiliary redox electrodes in each cell, to increase the available charge capacity and lengthen the operation duration between swaps. In their demonstration, Landman *et al.* placed four auxiliary redox electrodes in each cell, of which two electrodes were actively used during operation at one time and the other two were used at another time. Each pair of electrodes could supply enough charge for a single day (8 h) operation. Thus, using four auxiliary redox electrodes in each cell allowed to reduce the swapping frequency between the cells from daily swaps to once every two days. Ten unassisted water splitting cycles were demonstrated using solar-simulated light, with each cycle lasting for 8 h at an average solar to hydrogen (STH) conversion efficiency of 0.68%. In addition, operation under natural sunlight was also demonstrated.[34]

Another approach for decoupled PEC water splitting was presented by Bloor *et al.*, using phosphomolybdic acid (PMA, $(H_3O^+)[H_2PMo_{12}O_{40}]^-$ / $(H_3O^+)[H_4PMo_{12}O_{40}]^-$) as a soluble redox mediator and tungstic oxide (WO$_3$) photoanode in an acidic PEC cell.[33] Similarly to the previous system, the oxygen cell was the PEC cell, although in this case no additional bias was used during the oxygen generation step and therefore there was no need for a PV cell in tandem with the PEC cell as in the previous case. Under illumination, oxygen evolved at the WO$_3$ photoanode while the PMA was reduced at the auxiliary electrode according to:

$$[H_2PMo_{12}O_{40}]^- + 2e^- + 2H^+ \rightarrow [H_4PMo_{12}O_{40}]^- \quad \text{(Rxn 2)}$$

The reduced mediator, $[H_4PMo_{12}O_{40}]^-$, was then transferred to a second cell where it was re-oxidized at an auxiliary electrode that was biased at a potential of ~0.95 $V_{RHE}$ while hydrogen evolved at the cathode, completing the overall water splitting reaction.

In principle, the two cells could be operated concurrently and continuously, according to the operation scheme presented in Figure 4C. Unlike the Ni(OH)$_2$/NiOOH auxiliary redox electrodes that must be periodically swapped between the PEC cells in the solar field and the hydrogen generation unit, the PMA soluble redox mediator can simply be circulated between the oxygen and hydrogen cells. However, membranes must be installed in all cells to separate the catholyte and anolyte and prevent unintentional redox shuttling. Hence, while soluble redox mediators offer a potential reduction in operational complexity, solid auxiliary redox electrodes offer a potential alleviation of structural and maintenance complexities.

Finally, Li *et al.* demonstrated photocatalytic water splitting in an electrochemical – photocatalytic cycle using PMA as a soluble redox mediator and bismuth vanadate (BiVO$_4$) as a photocatalyst.[72] In the first step of their process, hydrogen evolved at the cathode while the PMA mediator was oxidized electrochemically at the auxiliary electrode by applying a potential of ~1.2 $V_{RHE}$ to the auxiliary electrode:

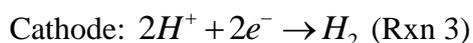

Cathode: $2H^+ + 2e^- \rightarrow H_2$  (Rxn 3)

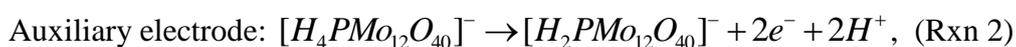

Auxiliary electrode: $[H_4PMo_{12}O_{40}]^- \rightarrow [H_2PMo_{12}O_{40}]^- + 2e^- + 2H^+$,  (Rxn 2)

In the next step, the oxidized [H$_2$PMo$_{12}$O$_{40}$]$^-$ redox mediator was photo-reduced in the presence of BiVO$_4$ particles. BiVO$_4$ is a semiconductor with a bandgap energy of 2.5 eV and a valence band edge located at ~7.27 eV below the vacuum level (*i.e.*, at a potential of ~2.82 $V_{RHE}$),[73] which is sufficient for water photo-oxidation.[74] Its conduction band edge is located at ~4.77 eV below the vacuum level (*i.e.*, at a potential of ~0.33 $V_{RHE}$),[73] which is not sufficient for hydrogen generation, but is sufficient for PMA photo-reduction. Thus, under illumination, the oxidized PMA was photo-reduced at the surface of the BiVO$_4$ photocatalyst while water was oxidized, evolving oxygen:

$$[H_2PMo_{12}O_{40}]^- + H_2O \xrightarrow{BiVO_4 + h\nu} [H_4PMo_{12}O_{40}]^- + \tfrac{1}{2}O_2 \quad \text{(Rxn 20)}.$$

Unlike the other two methods described above, the process proposed by Li *et al.* is photocatalytic rather than photoelectrochemical. Therefore, the oxygen evolution step requires only sunlight, without additional electrodes or wiring. Thus, in addition to the benefit of employing a soluble redox mediator that can be circulated between the hydrogen and oxygen cells, this process has the additional advantage of a simple photocatalytic oxygen cell without electrodes and without a membrane. However, the long-term stability of the system proposed by Li *et al.* is questionable, since BiVO$_4$ is known to be unstable in acidic solutions.[74]

## Summary

Decoupled water splitting offers several potential advantages over conventional water electrolysis. First and foremost, it offers separation of the hydrogen and oxygen evolution reactions over time and/or space. In conventional systems, including alkaline, PEM, AEM and solid oxide electrolyzers, H$_2$ and O$_2$ are produced concurrently in the same cell. A membrane or separator must therefore be installed between the cathode and anode to prevent the gases from mixing and recombining to water, reacting at the anode to form unstable reactive oxygen species that attack cell components, or creating combustible gas mixtures. These risks are reduced in decoupled water splitting, where hydrogen and oxygen are produced separately.

Partial-load conditions, inherent to intermittent renewable power sources such as solar and wind, give rise to an increased risk for hydrogen crossover through the membrane/separator in conventional electrolyzers. In contrast, decoupled water splitting systems could potentially operate under such conditions without risk of hydrogen cross permeation. Thus, decoupled water splitting presents an opportunity to safely integrate hydrogen generation from water with abundant renewable power sources such as solar and wind.

Essentially, a membrane or separator are no longer needed to separate the hydrogen and oxygen in decoupled systems with solid redox electrodes. Thus, solid redox electrodes enable membraneless operation, simplifying the electrolyzer construction and sealing, reducing maintenance requirements and potentially reducing material and construction costs. However, a membrane is still required in decoupled systems based on soluble redox mediators in order to prevent redox shuttling.

Despite separation of the primary electrodes, *i.e.*, the hydrogen evolving cathode and the oxygen evolving anode, there still remains a risk of parasitic HER and OER at the auxiliary electrodes. Ideally, the redox potential of the mediator should reside within the thermodynamic water stability region, *i.e.*, between 0 and 1.23 $V_{RHE}$ at 25°C. Inside this region, there is no risk of the hydrogen or oxygen evolution reactions competing with the mediator's reduction and oxidation reactions, respectively. However, some redox mediators can still facilitate decoupled water splitting even if their redox potentials lie outside of the thermodynamic water stability region, as long as they are within the *practical* stability region. The practical stability region is the potential range at which water is stable under the practical operating conditions. For example, while the OER standard potential is 1.23 $V_{RHE}$, significant oxygen evolution rates at most anode materials are not observed up to 1.6 $V_{RHE}$. The practical stability range is therefore between the onset potentials at which HER and OER occur at significant rates at the auxiliary electrodes. Thus, a redox mediator that undergoes oxidation above 1.23 $V_{RHE}$ but below the OER onset potential (under the same conditions) can still enable decoupled operation. The same is true for a redox mediator that is reduced below 0 $V_{RHE}$ but above the HER onset potential. However, at extreme states of charge, or if the redox reactions are carried out at high current densities, the risk for parasitic hydrogen and oxygen evolution increases. This issue becomes especially problematic if membraneless operation is desired. The risk of $H_2/O_2$ mixing and the limitations it imposes on the system operation, including limiting the SOC range and the current density, must therefore be taken into account.

Another advantage of decoupled operation is the possibility to operate under high pressure to produce pressurized hydrogen. Commercial electrolyzers typically operate at elevated pressures of 10 bar in alkaline electrolyzers, and 30 bar in PEM electrolyzers. However, many hydrogen-related applications require much higher pressures. Hydrogen storage tanks, for example, typically carry pressurized hydrogen at 350 or 700 bar. To reach such high pressures, the produced hydrogen gas must be compressed in an energy intensive and costly process. A much simpler and cheaper alternative is to perform high-pressure electrolysis in order to produce hydrogen at higher pressure. Such operation will reduce the compression ratio required from the compressor, thus reducing its cost. However, as the pressure increases, the hydrogen and oxygen bubbles become smaller and their solubility in the solution increases, resulting in cross-permeation through the membrane. Thus, decoupled water splitting, where hydrogen can be produced separately and without concurrent oxygen evolution, could enable the production of high-pressure hydrogen.

Decoupled electrolytic water splitting involves the introduction of another set of redox reactions, in addition to the HER and OER, namely, the oxidation and reduction of the redox mediator. The hydrogen cell requires a voltage equal to the difference between the redox mediator's oxidation potential and the HER potential. Similarly, the oxygen cell voltage is the potential difference between the OER and the redox mediator's reduction reaction. Therefore, whether the oxygen and hydrogen evolution reactions occur concurrently at different cells or in the same cell at alternating times, the overall voltage that is required to complete the water splitting reaction, $H_2O \rightarrow H_2 + \frac{1}{2}O_2$, is the sum of the oxygen cell and hydrogen cell voltages.

An ideal redox mediator would have negligible redox overpotentials, and the difference between its oxidation and reduction potentials would be close to zero. Otherwise, the overall voltage required for decoupled electrolytic water splitting would be greater than the voltage required for the equivalent coupled system (with only the primary anode and cathode) by an amount equal to the potential difference between the mediator's oxidation and reduction reactions. Consequently, the energy efficiency of decoupled electrolytic systems is necessarily lower than that of equivalent coupled systems. The overpotential losses over the mediator's redox reactions are influenced by several factors, such as the current density, the electrode's structure (in case of a solid redox electrode), the mediator's chemical composition, and the temperature and electrolyte concentration. Accordingly, the system and the redox mediator could be tailored to minimize these losses and balance the gains attributed to decoupled operation with the losses attributed to the increased power consumption.

Decoupled electrochemical – chemical water splitting schemes have the additional advantage of eliminating, at least in part, the voltage loss associated with cell separation. In these schemes, the water splitting reaction is divided into an electrochemical step, followed by a spontaneous chemical step. In the electrochemical step, oxygen (hydrogen) is evolved at the primary anode (cathode) while the redox mediator is reduced (oxidized) at the auxiliary cathode (anode). In the chemical step, the reduced (oxidized) redox mediator is then exposed to a catalyst or elevated temperature to accelerate its spontaneous chemical oxidation (reduction). This scheme involves only one electrochemical cell, with only two electrodes and two overpotentials, for the primary and auxiliary electrode reactions, instead of four overpotentials as in the electrolytic decoupling schemes.

As opposed to decoupled electrolytic schemes, where an ideal redox mediator's redox potentials lie between 0 and 1.23 $V_{RHE}$, an ideal redox mediator for decoupled electrochemical – chemical water splitting should have redox potentials slightly below or above 0 and 1.23 $V_{RHE}$, respectively. For example, if the redox mediator is reduced in the first step, its potential must be below 0 $V_{RHE}$ in order for it to be able to reduce water spontaneously. Therefore, such schemes replace either the HER or the OER with a redox reaction that necessarily has a lower reversible potential than the HER standard potential (0 $V_{RHE}$) or a higher potential than the OER standard potential (1.23 $V_{RHE}$). Nevertheless, both the HER and OER are seldom carried out at their standard potentials, and they may require expensive rare-Earth catalysts, as in PEM electrolysis. Thus, the overall voltage in electrochemical – chemical decoupling schemes can be *lower* than the corresponding coupled system voltage, or otherwise reduce the catalyst loading and cost. This is in addition to all of the previously mentioned advantages.

Electrochemical – chemical cycles take place in swing-mode operation, as opposed to continuous operation in conventional electrolysis and some decoupled electrolytic schemes. A soluble redox mediator is circulated from the electrochemical cell to the chemical cell, whereas a solid redox electrode remains static while different electrolytes are circulated through the cell. Since conventional electrolyzers also require electrolyte circulation to enable product separation, this does not appear to pose any specific limitation or disadvantage in decoupled compared to coupled systems.

In conclusion, after more than two centuries of coupled water electrolysis, decoupled water splitting shifts the paradigm away from simultaneous hydrogen and oxygen evolution and offers prospects for high-efficiency, robust, low-cost and high-pressure hydrogen production from renewable sources. This new and exciting field is emerging quickly, requiring vigorous research and development efforts, to identify more redox mediators, to evaluate their performance and stability, and to realize the full potential of decoupled water splitting for hydrogen production at industrial scales.

**Acknowledgment**

The authors would like to acknowledge support from the Israeli Center of Research Excellence on Solar Fuels, the Israeli Ministry of Energy, the European Research Council, the Nancy and Stephen Grand Technion Energy Program, the Adelis Foundation, and the Ed Satell lab for nitrogen hydrogen alternative fuels. G.S.G acknowledges the support of the Arturo Gruenbaum Chair in Material Engineering.